%% file: letter.tex
\documentclass[a4paper,12pt]{nature}
\usepackage[T1]{fontenc}
\usepackage[scaled]{helvet}
 
\usepackage{latexsym}
\usepackage[dvips]{epsfig}
\usepackage{float}

\usepackage{psfig}
\usepackage{amsmath}
\usepackage{epsfig}
\usepackage{fancybox}
\usepackage{graphicx}
\usepackage{amssymb}
\newcommand{\lya}{Ly$\alpha$}

\newcommand{\zz}{$z\sim$ 2}
\newcommand{\zzz}{$z\sim$ 3}
\newcommand{\zzzz}{$z\sim$ 4}

\def\simlt{\mathrel{\hbox{\rlap{\hbox{\lower4pt\hbox{$\sim$}}}
\hbox{$<$}}}}
\def\simgt{\mathrel{\hbox{\rlap{\hbox{\lower4pt\hbox{$\sim$}}}
\hbox{$>$}}}}

\title{Super-luminous supernovae at redshifts of 2.05 and 3.90}

\author{Jeff Cooke$^1$, Mark Sullivan$^2$, Avishay Gal-Yam$^3$,
Elizabeth J. Barton$^4$, Raymond G. Carlberg$^5$, Emma V.
Ryan-Weber$^1$, Chuck Horst$^6$, Yuuki Omori$^7$ \& C. Gonzalo
D\'{i}az$^1$}
 
\begin{document}
\spacing{1}
\maketitle
\thispagestyle{empty}
\begin{affiliations}
\item Centre for Astrophysics and Supercomputing, Swinburne University
  of Technology, PO Box 218, H30, Hawthorn, VIC, 3122, Australia
\item Department of Physics, University of Oxford, Denys
Wilkinson Building, Keble Road, Oxford OX1 3RH, UK
\item Benoziyo Center for Astrophysics, Weizmann Institute of Science,
76100 Rehovot, Israel
\item Center for Cosmology, Department of Physics and Astronomy,
University of California, Irvine, CA 92697, USA
\item Department of Astronomy and Astrophysics, University of
Toronto, Toronto, ON M5S 3H4, Canada
\item Department of Astronomy, San Diego State University, San Diego,
CA, 92182, USA
\item Department of Physics, McGill University, 3600 rue University,
Montreal, QC H2A 2T8, Canada
\end{affiliations}

\spacing{1}

\begin{abstract}

A rare class of `super-luminous' supernovae that are about ten or more
times more luminous at their peaks than other types of luminous
supernovae has recently been found at low to intermediate
redshifts.\cite{quimby11,gal-yam12} A small subset of these events
have luminosities that evolve slowly and result in radiated energies
of around 10$^{\mbox{\scriptsize 51}}$ ergs or more.  Therefore, they
are likely examples of `pair-instability' or `pulsational
pair-instability' supernovae with estimated progenitor masses of 100 -
250 times that of the Sun.\cite{rakavy67,barkat67,heger02} These
events are exceedingly rare at low redshift, but are expected to be
more common at high redshift because the mass distribution of the
earliest stars was probably skewed to high
values.\cite{larson98,bromm99} Here we report the detection of two
super-luminous supernovae, at redshifts of 2.05 and 3.90, that have
slowly evolving light curves.  We estimate the rate of events at
redshifts of 2 and 4 to be approximately ten times higher than the
rate at low redshift.  The extreme luminosities of super-luminous
supernovae extend the redshift limit for supernova detection using
present technology, previously 2.36 (ref.8), and provide a way of
investigating the deaths of the first generation of stars to form
after the Big Bang.

\end{abstract}

We search for high redshift super-luminous supernovae (SLSNe) by
modifying our image stacking and analysis technique\cite{cooke08} to
their expected light-curve evolution and high luminosities as recently
determined from theory and low-redshift observations.  Supernovae SN
2213-1745 and SN 1000+0216 were discovered when applying our modified
technique to the Canada-France-Hawaii Telescope Legacy Survey Deep
Fields.  The archival survey data provide deep, consistent photometry
of the host galaxies over $\sim$6 month seasons from 2003 to 2008 and
enable clean extraction of the supernova restframe far-ultraviolet
light.  Follow-up late-time spectroscopy of the supernovae and their
host galaxies was obtained using the 10-m Keck I telescope 5.2 and 3.8
yr ($\sim$626 and $\sim$286 d, restframe) after first detection for SN
2213-1745 and SN 1000+0216, respectively.  The deep Keck data reveal
redshifts of {\it z} = 2.0458 for SN 2213-1745 and {\it z} = 3.8993
for SN 1000+0216.  Details of the supernovae and their host galaxies
are listed in Table~\ref{table1}.

SN 2213-1745 is detected in the 2005 and 2006 seasonal (June -
November) stacked images for the field D4, and SN 1000+0216 is
detected in the 2006-2007 and 2007-2008 seasonal (November - June)
stacked images for field D2 (Figure 1).  Neither supernovae is
detected in the previous two seasons down to the stacked image
detection limits of m $\sim$ 26.5 mag (Supplementary Information,
section A).  The nightly images that comprise the seasonal stacks
first show the supernovae at the onset of the observing seasons.  The
lack of detection in previous seasons implies that the initial
outbursts occurred between November 2004 and June 2005 for SN
2213-1745 and between June and November 2006 for SN 1000+0216.  Both
supernovae evolve slowly and reach high peak luminosities of $\sim$0.5
- 1$\times$10$^{\mbox{\scriptsize 44}}$ erg s$^{\mbox{\scriptsize
-1}}$, uncorrected for host galaxy extinction.  The peak luminosities
correspond to far-ultraviolet absolute AB
magnitudes\cite{oke74,fukugita96} of M$_{\mbox{\scriptsize peak}}\sim$
-21 mag and thereby rule out all normal supernova subtypes.  SLSNe are
the only supernova type that can generate similar peak magnitudes and
light curve evolution (Figure 2).

Although active galactic nuclei (AGN), powered by the erratic
accretion of material onto supermassive black holes at the centres of
galaxies, can produce similar energies, both host galaxies exhibit a
single supernova-like outburst, present no activity during the
previous two years and have colours that are inconsistent with high
redshift AGN, including the eight spectroscopic AGN in our sample.
Moreover, the SN 2213-1745 host spectrum does not show signs of AGN
activity (Figure 3), and although the SN 1000+0216 host spectrum shows
narrow \lya\ emission, common to high redshift galaxies, no other
AGN-associated features are seen.  In contrast, the images reveal
zero, or very small, separations (projected in the plane of the sky)
from their host galaxy centroids (Supplementary Information, section
A).  We note that our technique is limited to detecting supernovae
within the small extent of the high-redshift host galaxies that
produces far-ultraviolet flux, and that far-ultraviolet light directly
traces regions of high star formation and does not necessarily provide
accurate galaxy centroids.  Finally, although the ultraviolet-optical
afterglows of long-duration gamma ray bursts can achieve luminosities
equivalent to, or greater than, the observed events, the bursts
typically reach their peak luminosities within a day and quickly
decay,\cite{kann11} usually following a power law of the form
t$^{\beta}$, with $\beta \sim$ -1.  This behaviour is inconsistent
with the observed slow rise and slow decay of the events discussed
here.

Recently, SLSNe have been classified into groups on the basis of their
photometric and spectroscopic properties\cite{gal-yam12}.  Those in
group SLSN-I show no evidence of hydrogen in their spectra, whereas
those in group SLSN-II are rich in hydrogen and include a subset
showing signs of interaction with circumstellar material. Finally,
those in group SLSN-R have light curves that evolve slowly, powered by
the radiative decay of $^{\mbox{\scriptsize 56}}$Ni.  SLSN-R events
are suspected to be pair-instability supernovae; the deaths of stars
with initial masses between 140 - 260 solar
masses\cite{rakavy67,barkat67,heger02}.  The physics of the
pair-instability process has been understood for many years, but the
extreme rarity of SLSN-R ($\sim$10 times less frequent than SLSN-I or
SLSN-II) has resulted in only one believable recorded event, SN
2007bi\cite{gal-yam09}, which is well studied up to late times, with a
few candidates being followed by ongoing low-redshift surveys.

In Figure 2, we compare the photometric evolution of the high-redshift
supernovae with lower-redshift SLSN data at similar
wavelengths.\cite{gezari09,barbary09,pastorello10,young10,chomiuk11}
All SLSNe, except for SLSN-R SN 2007bi which is believed to be powered
by $\sim$4 - 7 solar masses of $^{\mbox{\scriptsize 56}}$Ni, fade
significantly quicker than the two high-redshift events.  As far as
can be determined from the photometry, the evolution of SN 2213-1745
is consistent with SN 2007bi and, as a result, provides the first
far-ultraviolet data for an SLSN-R.  The close agreement suggests that
SN 2213-1745 may be powered by the radiative decay of a similar amount
of synthesised $^{\mbox{\scriptsize 56}}$Ni, and, along with an
integrated radiated luminosity of $\sim$10$^{\mbox{\scriptsize 51}}$
erg, implies a progenitor with an estimated initial mass of $\sim$250
solar masses.  We note that although SN 2213-1745 is shown to follow
closely the luminosity evolution of a radiation-hydrodynamics SLSN-R
simulation\cite{kasen11} for a progenitor star of similar mass, the
observed flux is higher and bluer than the model expectations
(Supplementary Information, section C).

SN 1000+0216 was observed over a shorter time period than was SN
2213-1745, as a result of time dilation.  The range of the light curve
sampled suggests a slower rise time than SN 2213-1745 but appears to
follow a similar fade rate from peak luminosity to $\sim$50 d later,
restframe, if we assume that the peak occurred during the gap in
coverage between the two detection seasons.  The peak far-ultraviolet
magnitude of SN 1000+0216 may exceed that possible for a
pair-instability supernova.  As a result, SN 1000+0216 may be an
example of a pulsational pair-instability supernova\cite{heger02} or a
SLSN-II similar to the low redshift SN 2006gy\cite{smith07,smith10},
which experience enhanced luminosity as a result of interaction with
previously expelled circumstellar material (Supplementary Information,
sections D \& F).  The high luminosity of SN 1000+0216 classifies it
as a SLSN but, because of its limited photometric coverage and low
signal-to-noise host spectrum, its subclassification remains
uncertain.

Our programme searches for {\it z} $\gtrsim$ 2 supernovae by
monitoring the well-studied population of Lyman break
galaxies\cite{steidel99,steidel03,steidel04} over several well-defined
volumes.  Because Lyman break galaxies comprise the bulk of galaxies
at high redshift, where star formation rates are higher, normal
populations of short-lived, massive stars are more common at {\it z}
$\gtrsim$ 2 than locally.  From the specifics of our survey, the two
SLSN detections imply a rough volumetric high-redshift rate of a few
$\times$ 10$^{\mbox{\scriptsize -7}}$ {\it h}$_{\mbox{\scriptsize
71}}^{\mbox{\scriptsize 3}}$ Mpc$^{\mbox{\scriptsize -3}}$
yr$^{\mbox{\scriptsize -1}}$ at {\it z} = 2.2 $\pm$0.3 and {\it z} =
4.1 $\pm$0.3 (mean $\pm$1$\sigma$)(Supplementary Information, section
B).  After correcting for the increase in the cosmic star formation
rate from low to high redshift, the SLSN rate remains $\sim$10 times
higher than that estimated at low-redshift\cite{quimby11}, but we
caution that our rate estimate is poorly constrained because it is
derived from only two events.  The far-ultraviolet photometry confirms
that the two supernovae are strong sources of escaping high-energy
photons.  However, far-ultraviolet light is highly susceptible to
metal-line absorption and local and global dust extinction that may
have a greater effect on other high redshift SLSNe and cause them to
fall below our detection threshold.  As a result, our estimated rate
is a lower limit and implies that the discrepancy between low and high
redshift may be even greater.

The detection of SLSNe at {\it z} $>$ 2 presents the possibility of
finding the explosions of Population III stars, the first stars to
form after the Big Bang.  Population III stars are predicted to exist
a redshifts as low as {\it z} $\sim$ 2 (refs 25-28) and have mass
distributions skewed towards high masses\cite{larson98,bromm99}.  On
the basis of our late-time spectroscopy, the supernovae presented here
are unlikely to be from the first generation stars (Supplementary
Information, section E).  Deep spectroscopy of future supernovae
obtained near maximum brightness, and their use as sightline probes,
offers a means to help distinguish which events formed in regions with
essentially no enrichment in elements heavier than helium, and thus
likely had Population III progenitors.

\begin{addendum}

\item We acknowledge support from ARC, NSERC and the Royal Society.
AGY acknowledges support by ISF, GIF, and Minerva grants, an ARCHES
award, and the Lord Seiff of Brimpton Fund.  Supported in part by a
grant from the ANSTO AMRFP.  The access to major research facilities
program is supported by the Commonwealth of Australia under the
International Science Linkages program.  The results presented here
are based on observations obtained with MegaPrime/MegaCam, a joint
project of CFHT and CEA/DAPNIA, at the Canada-France-Hawaii Telescope
(CFHT) which is operated by the National Research Council (NRC) of
Canada, the Institut National des Sciences de l'Univers of the Centre
National de la Recherche Scientifique (CNRS) of France, and the
University of Hawaii. This work is based in part on data products
produced at TERAPIX and the Canadian Astronomy Data Centre as part of
the Canada-France-Hawaii Telescope Legacy Survey, a collaborative
project of NRC and CNRS.  We used data products from the Canadian
Astronomy Data Centre as part of the CFHT Legacy Survey.  The
spectroscopic data presented herein were obtained at the W.M. Keck
Observatory, which is operated as a scientific partnership among the
California Institute of Technology, the University of California and
the National Aeronautics and Space Administration.  The Observatory
was made possible by the generous financial support of the W.M. Keck
Foundation.  

\end{addendum}

\clearpage

\begin{table}
\centering
{\footnotesize
\begin{tabular}{lcccccc}
\hline Supernova & R.A. & Dec. & Host & SN peak & Detection & Redshift\\
& (J2000) & (J2000) & M$_{\mbox{\scriptsize FUV}}$ & M$_{\mbox{\scriptsize FUV}}$ & dates & {\it (z)}\\
\hline
SN 2213-1745 & 22 13 39.970 & -17 45 24.486 & -21.38$\pm$0.03 & 
-21.2$\pm$0.2 & Jun 2005 - Nov 2006 & 2.0458$\pm$0.0005\\
SN 1000+0216 & 10 00 05.862 & +02 16 23.621 & -21.20$\pm$0.04 & 
-21.6$\pm$0.2 & Nov 2006 - Jun 2008 & 3.8993$\pm$0.0074\\
\hline
\hline
\end{tabular}}
\vspace{0.2cm}

\caption{Supernova and host galaxy details.  The far-ultraviolet
absolute magnitudes, M$_{\mbox{\scriptsize FUV}}$ are derived using
the conventional relationship M$_{\mbox{\scriptsize FUV}}$ =
m$_{\mbox{\scriptsize AB}}$ - 5$\cdot$log$_{\mbox{\scriptsize
10}}$({\it D}$_{\mbox{\scriptsize L}}$({\it z})/10pc) +
2.5$\cdot$log$_{\mbox{\scriptsize 10}}$(1+{\it z}), where
m$_{\mbox{\scriptsize AB}}$ is the observed redshifted far-ultraviolet
AB magnitude\cite{oke74,fukugita96} and {\it D}$_{\mbox{\scriptsize
L}}$ is the luminosity distance, adopting a standard {\it H}$_{\mbox
\scriptsize 0}$ = 71, $\Omega_\Lambda$ = 0.73, $\Omega_M$ = 0.27
cosmology.  The supernova absolute magnitudes are corrected for
extinction by the Milky Way Galaxy\cite{schlegel98} (0.04 - 0.11 mag)
but not corrected for supernova host galaxy extinction.  }

\label{table1}
\end{table}

\clearpage

\begin{figure}
\centerline{\psfig{figure=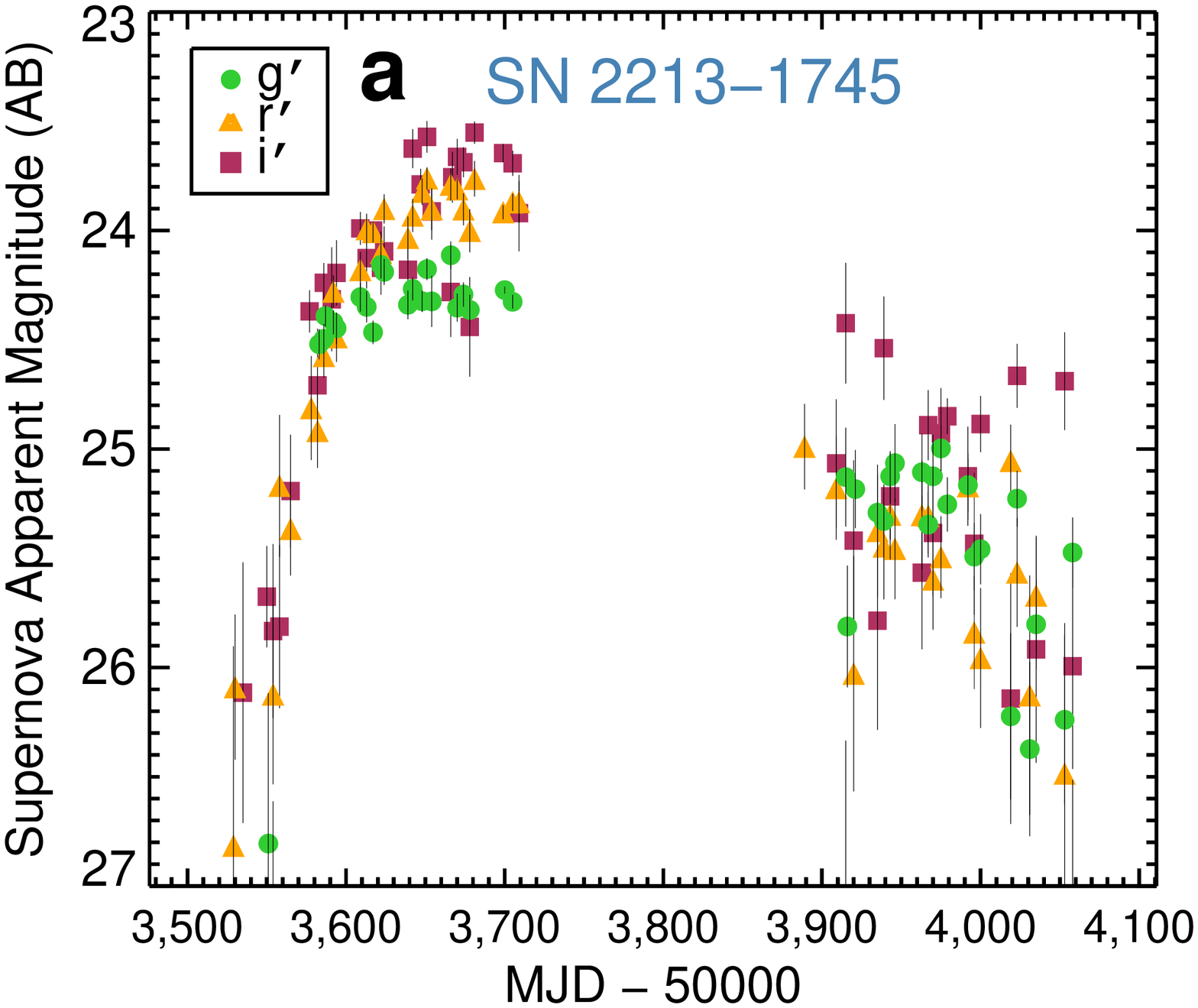,angle=90,height=8.5cm,width=10.5cm}}
\centerline{\psfig{figure=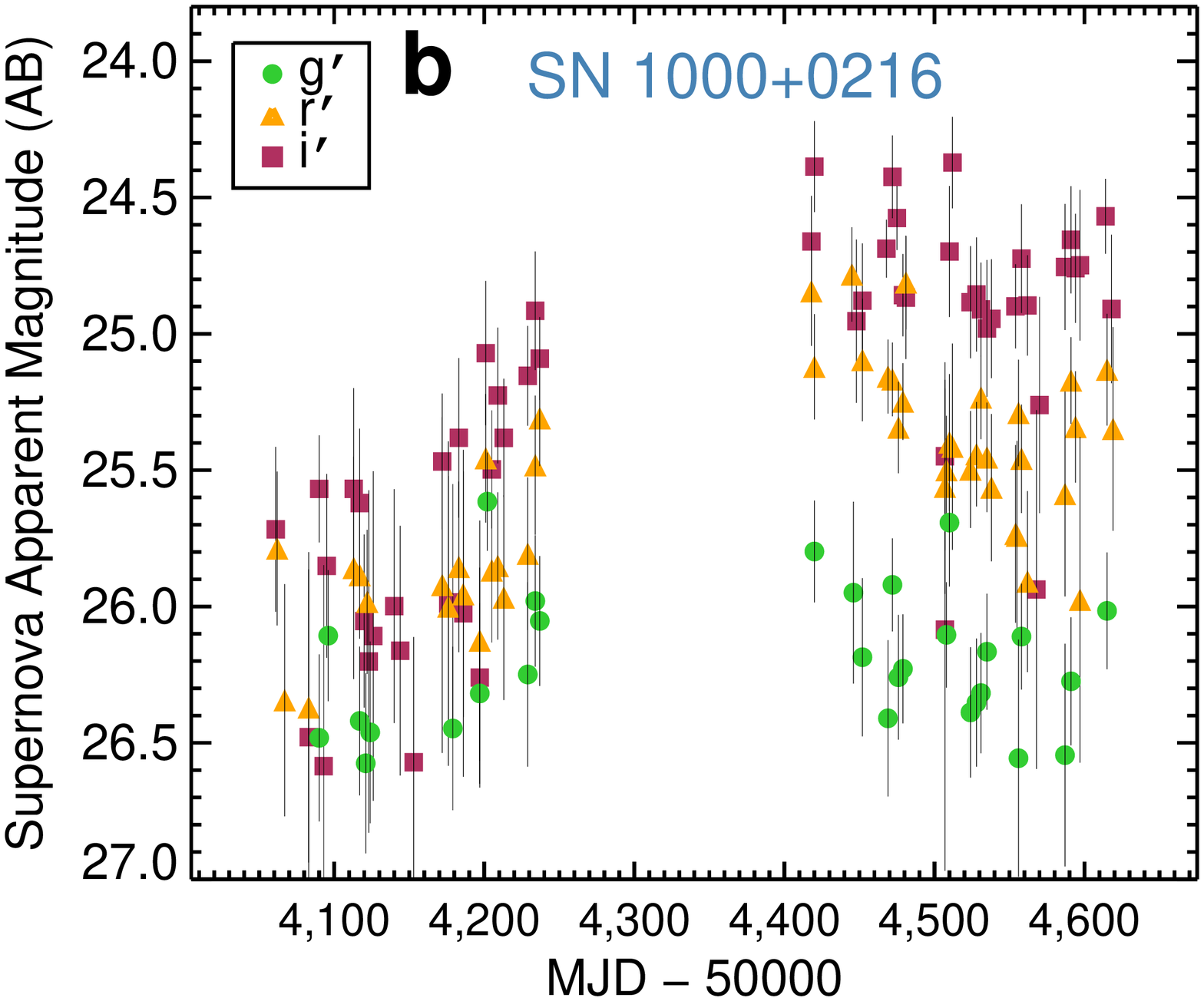,angle=90,height=8.5cm,width=10.5cm}}

\caption{\small Light curves of high-redshift supernovae.  Plotted as
functions of time (MJD, modified Julian date) are the observed
supernova {\it g'-}, {\it r'-}, and {\it i'-}band magnitudes and
1$\sigma$ errors detected in nightly stacked images after host galaxy
flux subtraction.  The high quality images used in our study provide
consistent and accurate photometry of the supernova host galaxies for
$\sim$6 months per year for four years (Supplementary Information,
section A). {\bf (a)} Magnitudes for SN 2213-1745 ({\it z} = 2.05)
detected in deep field D4 during the observing seasons June - November
2005 and June - November 2006.  The optical {\it g'-}, {\it r'-}, and
{\it i'-}band data correspond to restframe far-ultraviolet continuum
wavelengths of $\sim$1,600\AA, $\sim$2,050\AA, and $\sim$2,530\AA,
respectively.  {\bf (b)} Magnitudes for SN 1000-0216 ({\it z} = 3.90)
detected in deep field D2 during the observing seasons November 2006 -
June 2007 and November 2007 - June 2008.  Here, the {\it g', r',} and
{\it i'} filters correspond to restframe wavelengths of
$\sim$1,000\AA, $\sim$1,300\AA, and $\sim$1,600\AA, respectively.  As
a result, only the {\it i'} filter samples the far-ultraviolet
continuum exclusively.  The {\it r'} filter probes the continuum and
the Lyman $\alpha$ (\lya) feature and samples the decrement in flux
shortward of \lya\ caused by optically thick systems at lower redshift
in the line of sight, termed the \lya\ forest.  The {\it g'} filter
completely samples the spectral region of the \lya\ forest to just
below the Lyman limit.}

\end{figure}

\begin{figure}
\centerline{\psfig{figure=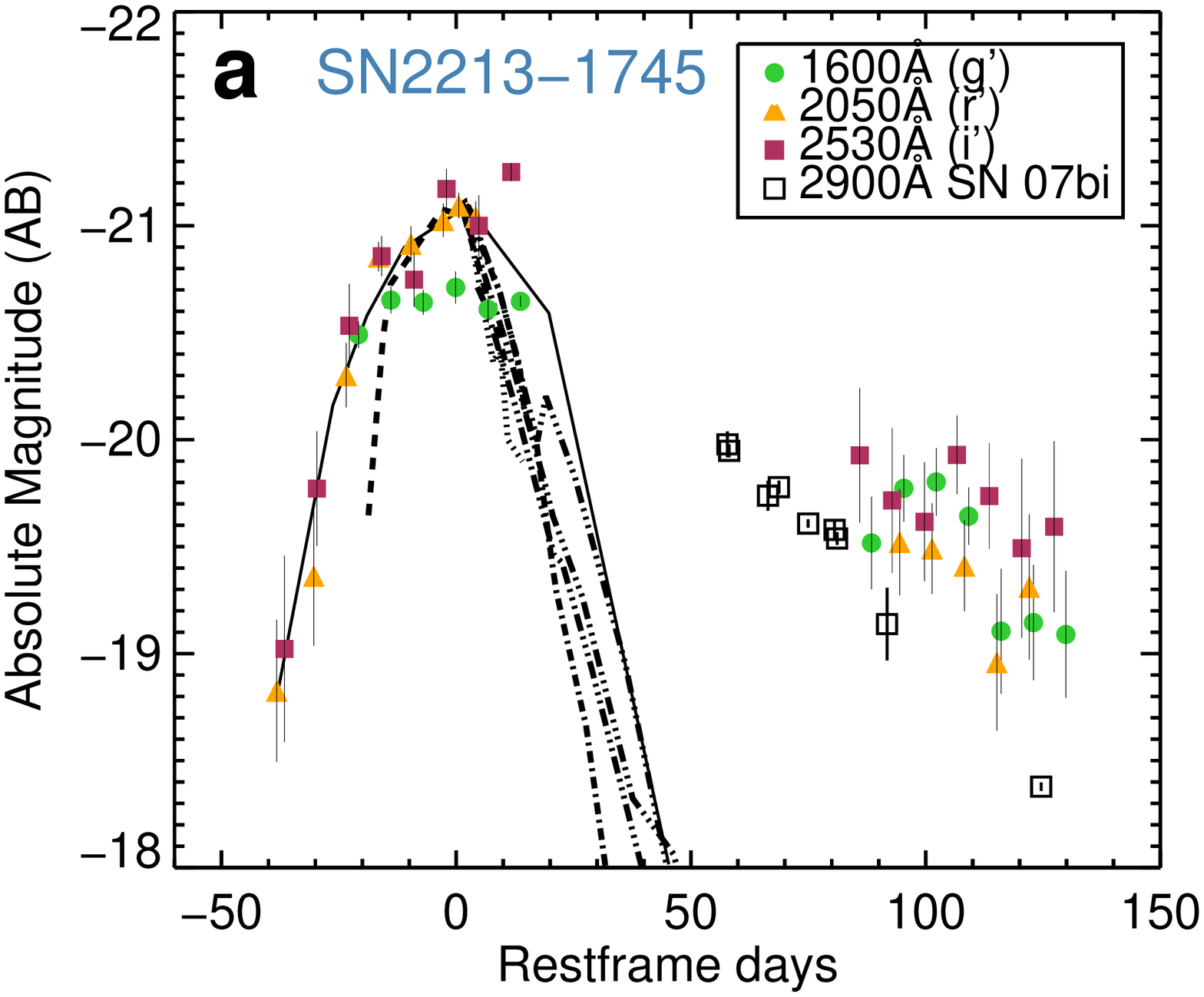,angle=90,height=8.0cm,width=10.5cm}}
\centerline{\psfig{figure=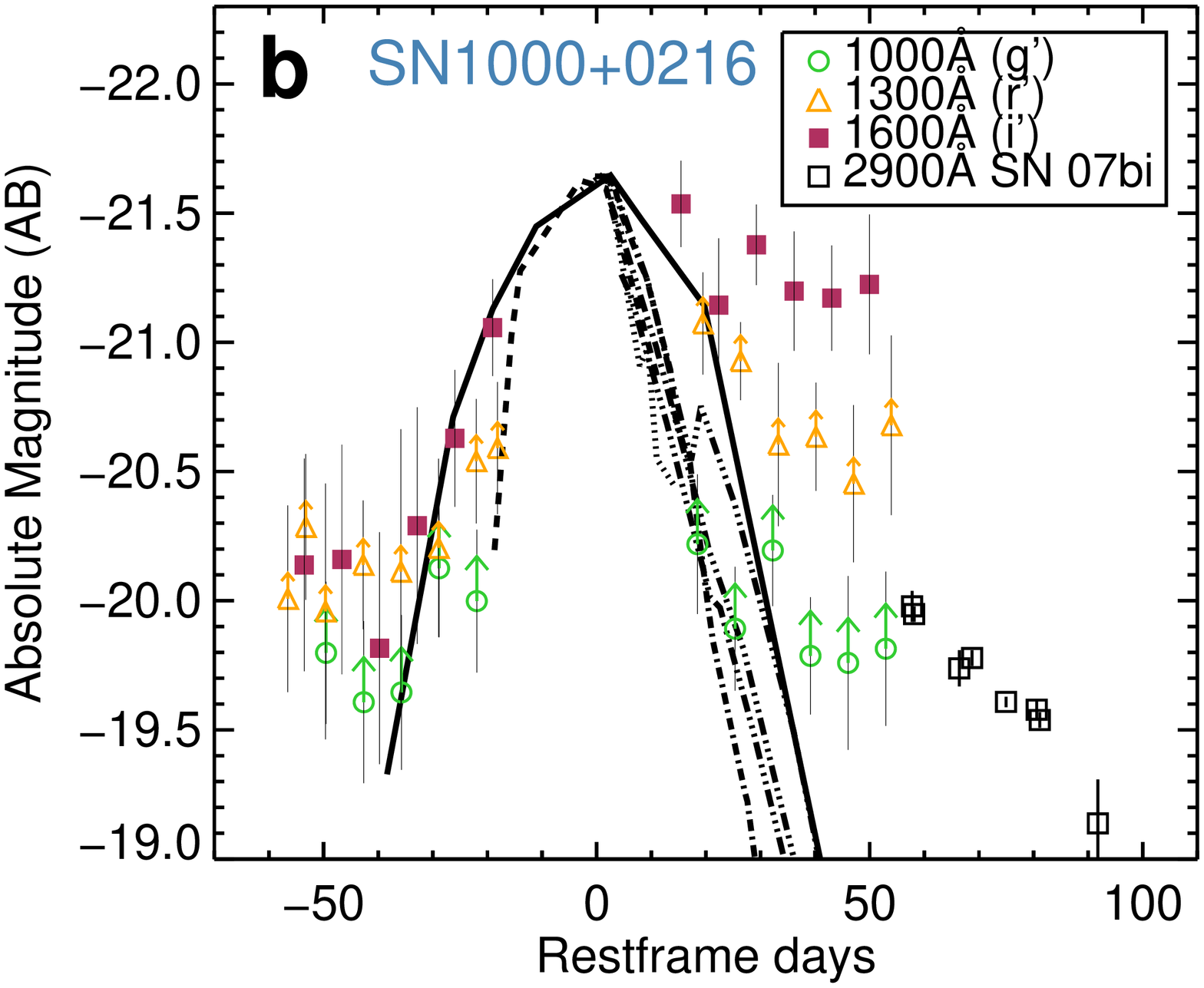,angle=90,height=8.0cm,width=10.5cm}}

\caption{\small Absolute magnitude light curves of super-luminous supernovae.
{\bf (a)} The rise and decay of the luminosity of SN 2213-1745 in the
supernova rest frame, binned in 7-day intervals.  At {\it z} = 2.05,
the {\it g',r'} and {\it i'} filters probe effective wavelengths of
1,600\AA, 2,050\AA, and 2,530\AA, respectively.  For comparison of
light curve evolution, the photometry of six lower-redshift
super-luminous supernovae observed at similar, but typically longer,
wavelengths is overlaid and scaled by 0 - 1.2 mags to match the
estimated peak light of the data: SCP 06F6\cite{barbary09} (SLSN-I,
3,500\AA, solid curve), SN 2010gx\cite{pastorello10} (SLSN-I,
2,900\AA, dotted curve), PS1-10awh\cite{chomiuk11} (SLSN-I, 2,500\AA,
dashed curve), PS1-10ky\cite{chomiuk11} (SLSN-I, 2,500\AA, dot-dashed
curve), and SN 2008es\cite{gezari09} (SLSN-II, 1,700 - 2,200\AA,
dot-dot-dashed curves).  Also shown is type SLSN-R SN
2007bi\cite{young10} with zero magnitude offset.  We note that the SN
2007bi data are based on the known peak magnitude from r-band
photometry acquired from just before peak to $>$300 days after
peak.\cite{gal-yam09} Error bars on all points represent 1$\sigma$.
{\bf (b)} Same as (a) but for SN 1000+0216 at {\it z} = 3.90.  Here,
the lower redshift super-luminous supernovae are matched to an
estimated peak in the {\it i'} filter (1,600\AA) which is the only
filter that samples the far-ultraviolet continuum of the supernova
exclusively.  This arbitrary peak for SN 1000+0216 results in a -0.1
to -0.6 mag offset for the lower redshift SLSNe-I and SLSN-II (as in
(a), SN 2007bi is not offset).  As described in Figure 1, the {\it g'}
(1,000\AA) and {\it r'} (1,300\AA) filters partially or completely
probe the spectral region of the \lya\ forest.  The upward arrows
indicate that the true flux in these filters is higher than
indicated.}

\end{figure}

\begin{figure}
\centerline{\psfig{figure=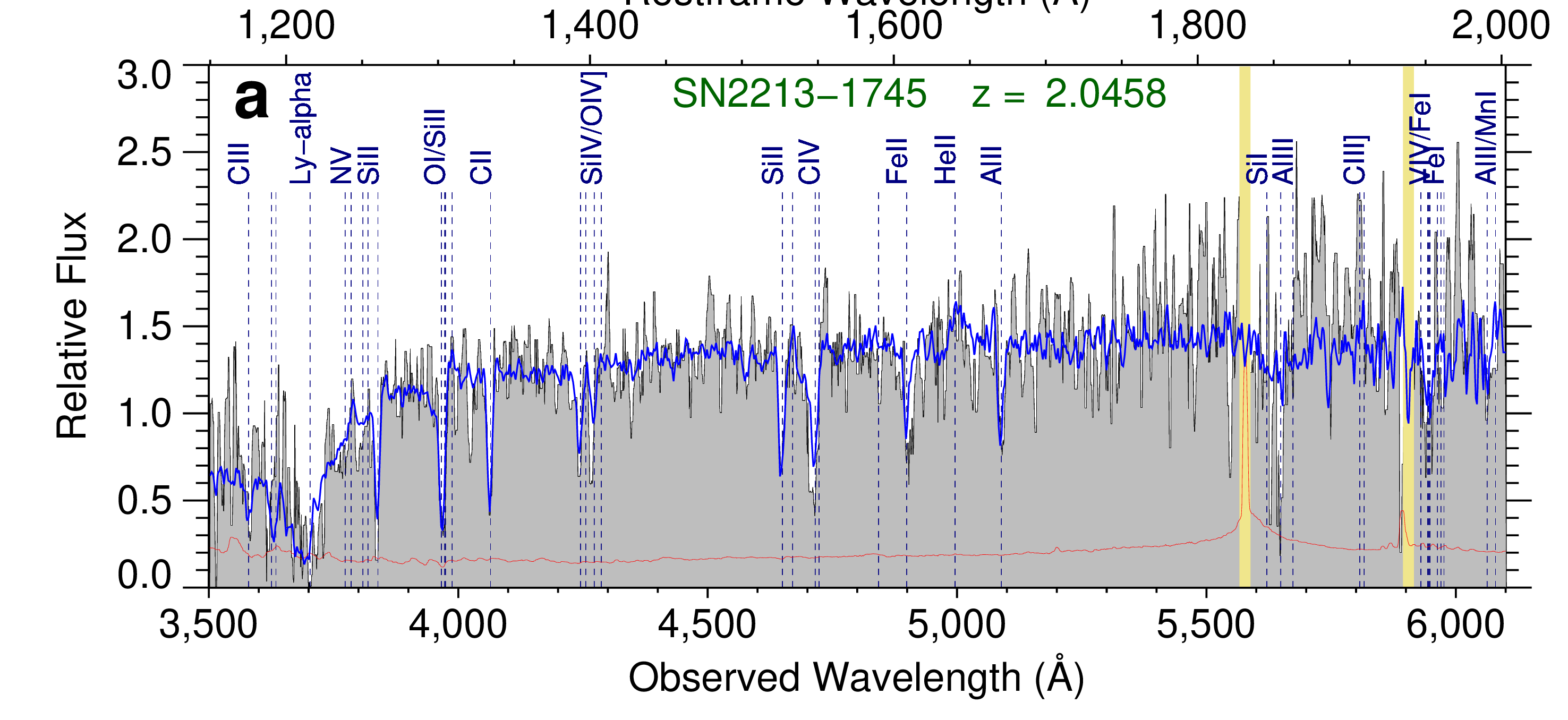,angle=90,height=7.5cm,width=16.0cm}}
\vspace{0.6cm}
\centerline{\psfig{figure=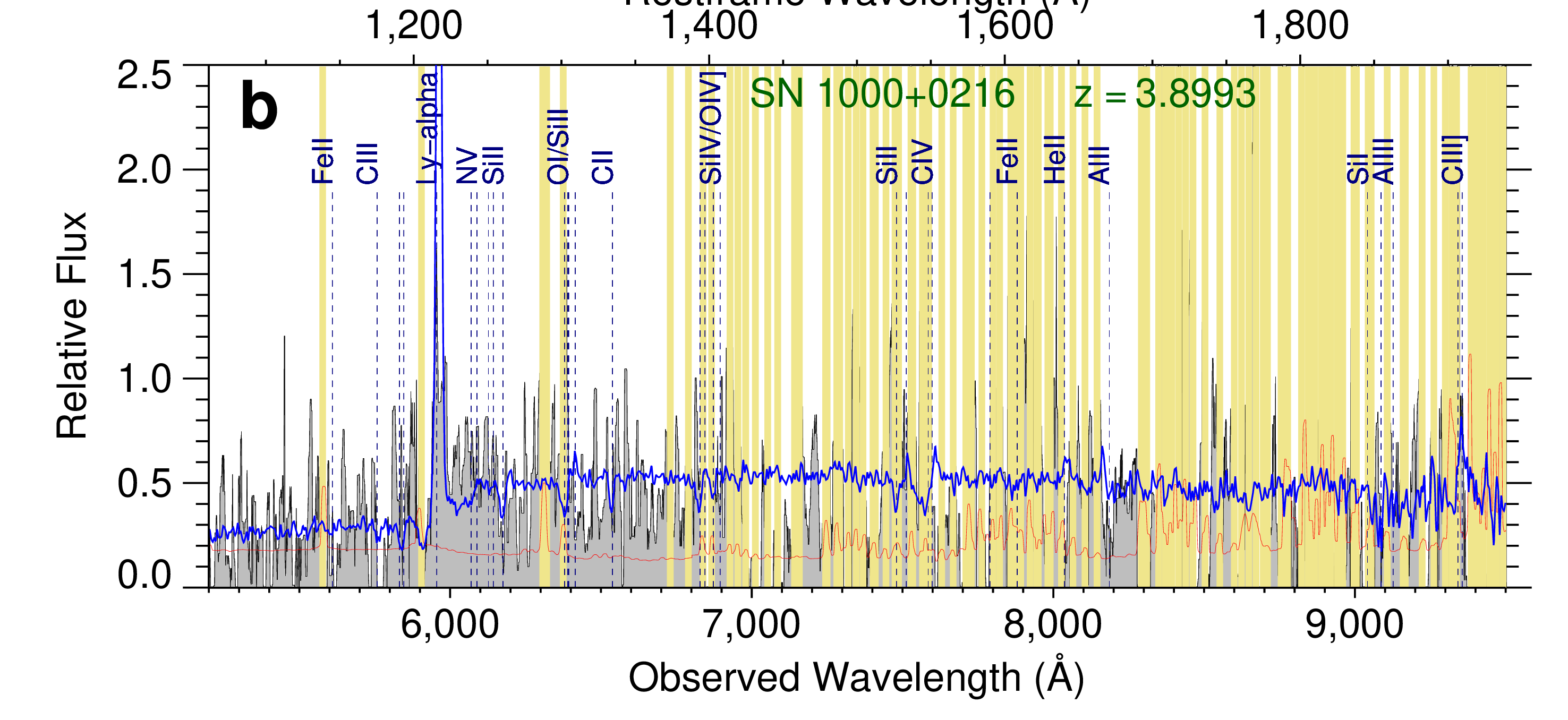,angle=90,height=7.5cm,width=16.0cm}}

\caption{\small Late-time spectra of the supernovae and host galaxies.  The
spectra are shown as the dark grey filled regionS.  Although Lyman
break galaxies are extremely distant, they have strong absorption
features and a prominent \lya\ feature at 1216\AA, seen in absorption
and/or emission, which enable reliable identification.  Labelled
vertical dashed lines indicate typical atomic transitions seen in
absorption.  Additional absorption features may be present as a result
of lower-redshift systems in the line of sight.  Thick vertical
light-gray lines mark the positions of bright night-sky emission lines
that are difficult to remove cleanly from faint spectra.  {\bf (a)}
Spectrum of SN 2213-1745 obtained 626 days (restframe) after first
detection.  The 1$\sigma$ per pixel error array is shown by the thin
black curve.  A composite template of $\sim$200 similar spectra from
Lyman break galaxies (thick black curve) is overlaid and provides a
secure galaxy identification and redshift.  The spectrum shows no
evidence of AGN activity such as strong \lya\ 1,216\AA, N\textsc{v}
1,240\AA, or C\textsc{iv} 1,550\AA\ emission.  {\bf (b)} Spectrum of
SN 1000+0216 obtained 286 days (restframe) after first detection
plotted as in (a).  The poorer observing conditions and fainter nature
of this object resulted in a spectrum with a lower signal-to-noise
ratio in more difficulty subtracting bright night-sky lines.  However,
a confident host galaxy identification and redshift is achieved in
part on the basis of the significant narrow \lya\ emission and
decrement in continuum flux shortward of \lya\ as a result of the
\lya\ forest.  Although the host galaxy exhibits narrow \lya\ in
emission (observed in $>$ 50\% of Lyman break galaxies), associated
AGN features, such as N\textsc{v} 1,240\AA\ and C\textsc{iv} 1,550\AA,
are not seen at any significance.  Flux correction between $\sim$6,300
- 6,800\AA\ is under-represented and less reliable (Supplementary
Information, section A). }

\end{figure}

\clearpage
\include{si}

\end{document}

%% file: si.tex
\textheight 10.0in
\textwidth 6.5in
\topmargin  -0.15in
\oddsidemargin -0.1in
\evensidemargin 0.0in

\bibliographystyle{naturemag}
\pagestyle{empty} 

\appendix
\numberwithin{figure}{section}
\numberwithin{table}{section}
\markright{Super-luminous supernovae at redshifts of 2.05 and 3.90}

\Large\noindent{\bf Supplementary Information} 
\normalsize 
\spacing{1}

\noindent Our programme searches deep, yearly stacked images for
far-ultraviolet (FUV) luminous supernovae by comparing the
year-to-year restframe FUV flux of {\it z} $>$ 2 Lyman break galaxies
(LBGs).  The bulk of all FUV-luminous supernovae observed to date are
type IIn supernovae and, more recently, a class of super-luminous
supernovae (SLSNe) in which both are believed to result from the
deaths of massive stars.  Type IIn supernovae are defined by the
narrow emission features in their spectra.  Interaction with dense,
circumstellar material results in long-lived, extremely luminous,
narrow emission lines that typically remain strong for $\sim$1 -
2$\cdot$(1+{\it z}) years after outburst.  As a result, we follow-up
supernova candidates with deep spectroscopy to identify the host
galaxy redshifts and to search for late-time supernova emission.


\section{Observations}
\label{supp_obs}

\subsection{Photometry} 
\label{phot}
 
We use high-quality (seeing FWHM $<$0.75 arcsec) images of the
Canada-France-Hawaii Telescope Legacy Survey (CFHTLS) Deep fields for
our photometry {\it (Information for the CFHTLS survey and publicly
available data products can be found at a number of associated
websites including http://www.cfht.hawaii.edu/Science/CFHLS/)}.  LBGs
are colour-selected in five bandpasses ({\it u$^*$g'r'i'z'}) in
stacked images comprised of data taken over four years
(m$_{\mbox{\scriptsize lim}}\sim$27) with consistent {\it i'}-band
imaging$^{\mbox{\scriptsize 8}}$.  Yearly stacked {\it g'r'i'} images
(m$_{\mbox{\scriptsize lim}}\sim$26.5) are searched for supernova-like
events and consist of multiple frames taken per night over $\sim$25 -
30 nights during $\sim$6 month seasons in which the fields are
observable.  We focus on the {\it g'r'i'} yearly stacked images
because of their superior depth and even cadence as compared to the
{\it u$^*$} and {\it z'}-bands.  In addition, the Lyman limit is
redshifted to the {\it u$^*$} and {\it g'} bands for \zzz\ and \zzzz\
objects, respectively, and results in weak to essentially no
observable flux depending on object redshift.

\begin{figure}[h!]
\begin{center}
\scalebox{0.34}[0.33]{\rotatebox{90}{\includegraphics{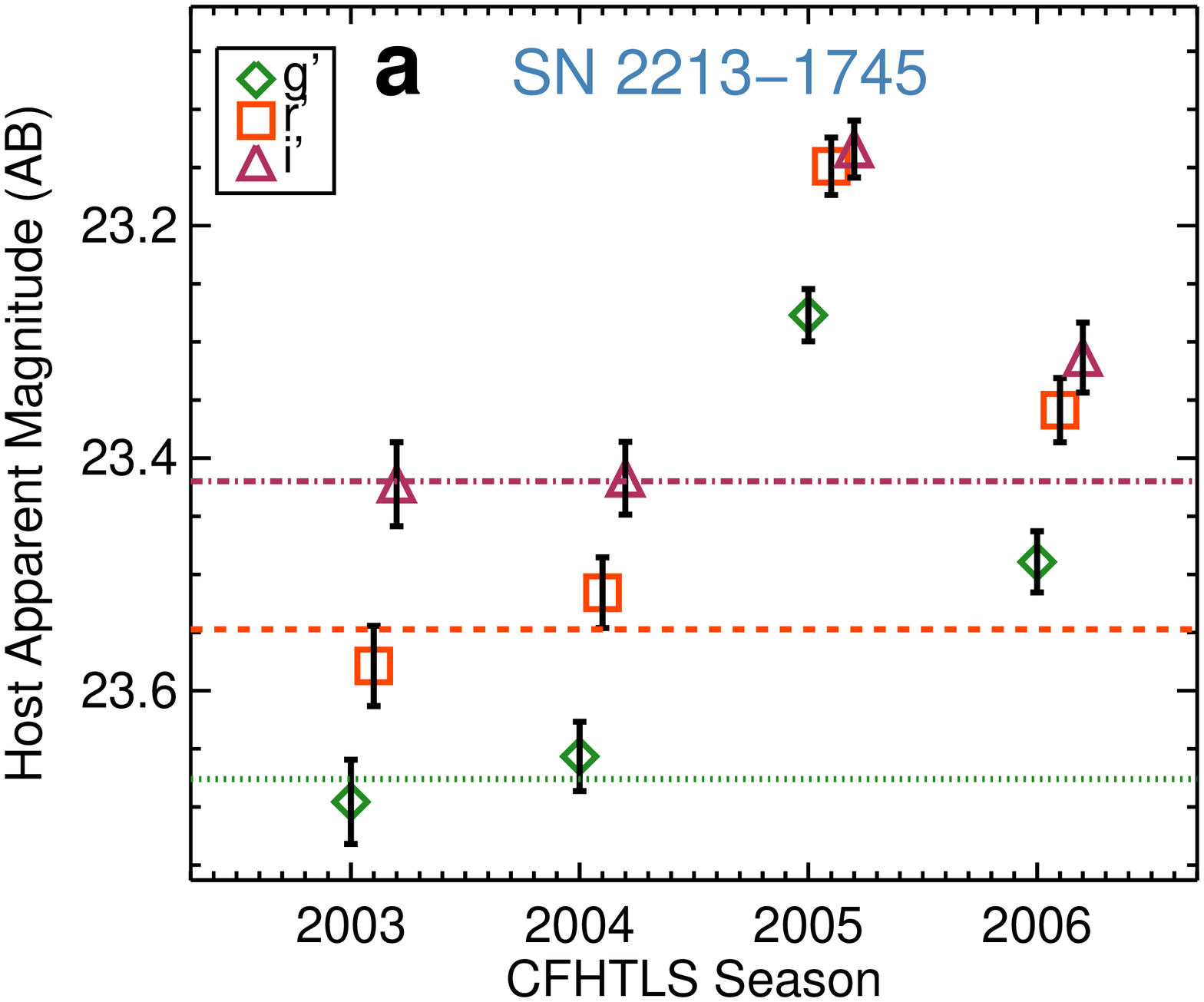}}}
\scalebox{0.34}[0.33]{\rotatebox{90}{\includegraphics{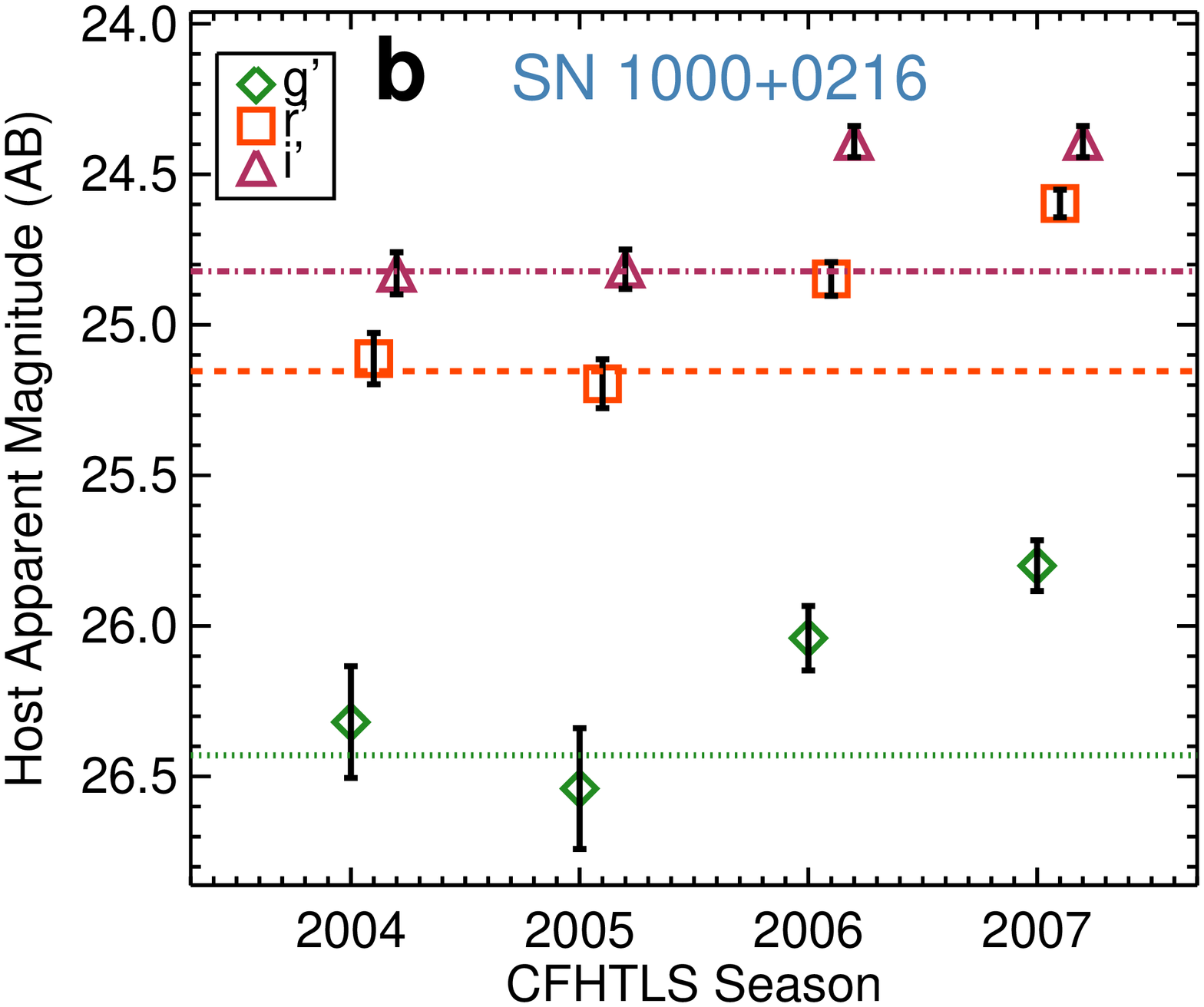}}}
\scalebox{0.34}[0.33]{\rotatebox{90}{\includegraphics{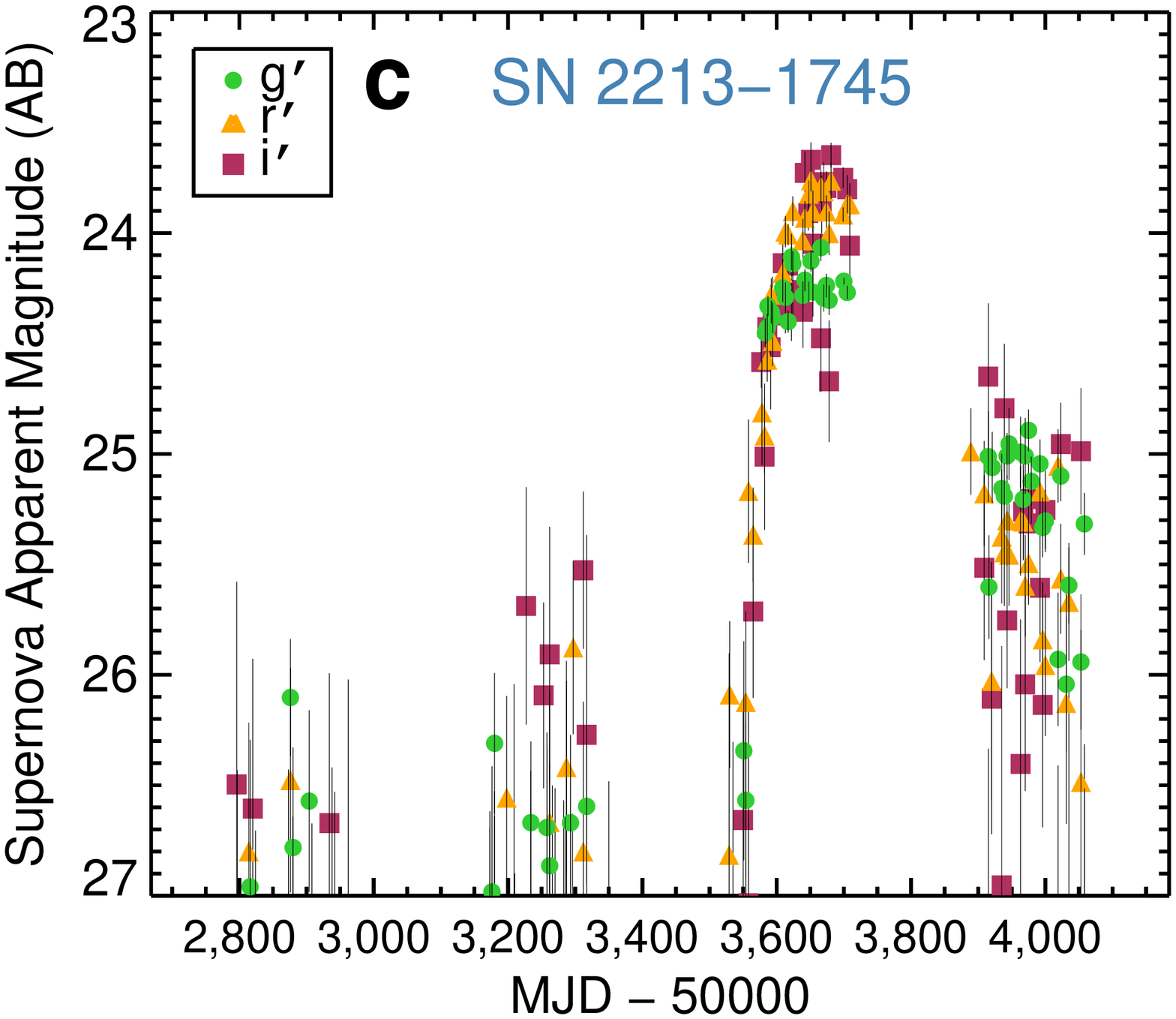}}}
\scalebox{0.34}[0.33]{\rotatebox{90}{\includegraphics{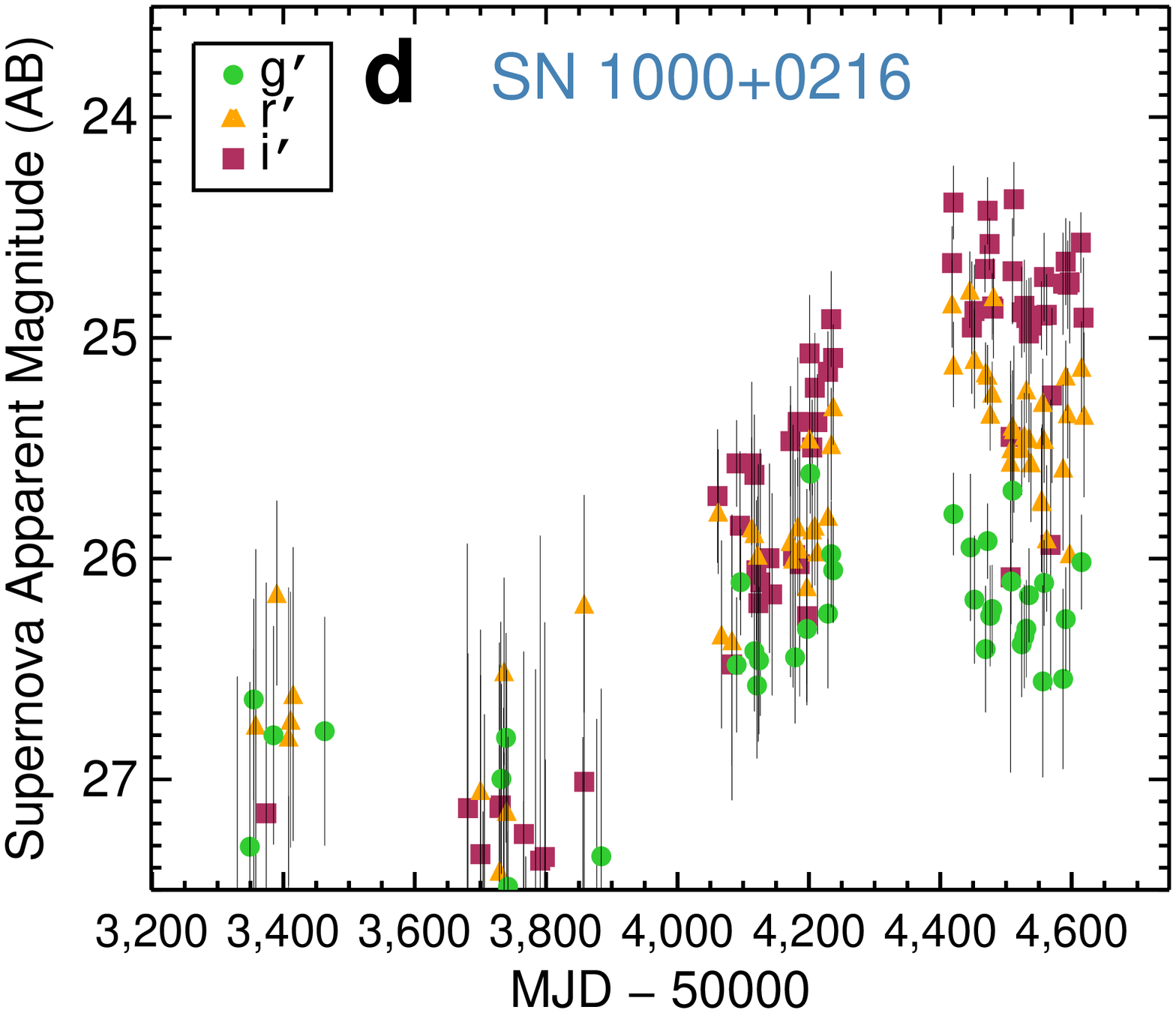}}}

\caption {Supernova and host galaxy magnitudes.  {\bf (a)} SN
2213-1745 host Lyman break galaxy magnitudes monitored over four
years.  The {\it g'r'i'} magnitudes and 1$\sigma$ errors from the
yearly seasonal stacked images are shown.  The dotted, dashed, and
dot-dashed lines mark the intrinsic host magnitude as determined from
the non-detection seasons in {\it g',r',} and {\it i'}, respectively.
Typically, FUV luminous supernovae at \zz\ are detectable in only one
year, however, SN2213-1745 was energetic enough for a significant
detection in the subsequent (2006) observing season.  {\bf (b)}
Similar to panel (a), but for SN1000+0216 host Lyman break galaxy.
The high luminosity and high redshift of SN1000+0216 resulted in a
detection in 2006 and 2007.  {\bf (c)} Nightly image stacks tracing
the evolution of SN2213-1745 over four years after subtraction of the
host galaxy flux. {\bf (d)} Similar to panel (c), but for SN
1000+0216.}

\label{yearly}
\end{center} 
\end{figure}

\noindent The rise and decay times for typical FUV luminous supernovae
at \zz\ - 4 are such that, including time dilation, they are only
detectable in the integrated flux for a single $\sim$6 month observing
season (or yearly stacked image).  The supernovae fade sufficiently to
evade detection in the following observing season due, in part, to the
$\sim$6 months gap between observations.  The intense luminosity and
slow rise and decay of SLSNe, however, result in high redshift
detections that can straddle two (or three) observing seasons.  The
integrated yearly host galaxy magnitudes for SN2213-1745 and
SN1000+0216 and supernova magnitudes from the nightly stacked images
obtained over four years are shown in Figure~\ref{yearly}.
SN2213-1745 is not detected in the integrated flux of the host galaxy
in years 2003 and 2004 and is detected at $\sim$3 - 12$\sigma$ during
2005 and 2006.  Similarly, SN1000+0216 is not detected in 2004/2005
and 2005/2006 and is detected at $\sim$2 - 10$\sigma$ in 2006/2007 and
2007/2008.  The $\sim$6 month observing seasons for the CFHTLS D2
field, in which SN1000+0216 is located, spanned November in one year
through June in the next.

\subsection{Spectroscopy}
\label{spec} 

Spectroscopic observations for SN2213-1745 and its host LBG were
acquired on 03 September 2010 using the dual-arm Low-Resolution
Imaging Spectrometer (LRIS)$^{\mbox{\scriptsize 24,}}$\cite{oke95} and
atmospheric dispersion corrector on the Keck I telescope.  The data
were obtained using a D560 dichroic and the 400 line
mm$^{\mbox{\scriptsize -1}}$ grism blazed at 3,400\AA~on the blue arm
and the 400 line mm$^{\mbox{\scriptsize -1}}$ grating blazed at
8,500\AA~on the red arm.  We obtained 4 $\times$ 1,200s exposures
using a 1.0 arcsec slit under good conditions and $\sim$0.8 arcsec
seeing FWHM.  At the time of the observations, the red arm electronics
were experiencing significant problems that rendered three quarters of
the red arm CCD area suspect for scientific use.  Although the
spectrum was placed on the area of the CCD that was `good', we
consider any result at wavelengths greater than 5,600\AA\
($\sim$1,850\AA, restframe) tentative.  The intrinsic luminosity of
the SN2213-1745 host LBG afforded a final per pixel signal-to-noise
ratio of $\sim$5 - 10.  Observations for SN1000+0216 and its host LBG
were acquired on 10 March 2011 with LRIS using the D680 dichroic, 300
line mm$^{\mbox{\scriptsize -1}}$ grism blazed at 5,000\AA~(blue arm),
and 400 line mm$^{\mbox{\scriptsize -1}}$ grating blazed at
8,500\AA~(red arm).  We obtained 6 $\times$ 1,200s exposures using a
1.5 arcsec slit on a multi-object mask under moderate conditions, with
$\sim$1.1 arcsec seeing FWHM and light cirrus.  Because no
order-blocking filter is available for LRIS on the blue arm to
eliminate second-order flux from standard star observations, the flux
corrections to the spectrum from $\sim$6,300 - 6,800\AA~are
over-corrected (i.e., weaker than the true value) by a small factor.
Both spectra are presented in Figure 3 in the main Letter and the
spectrum of SN2213-1745 is reviewed again in Figure~\ref{hostspec}.

\subsection{Supernova location}
\label{centroids}

\begin{figure}[h!]
\begin{center}
\scalebox{0.25}[0.25]{\includegraphics{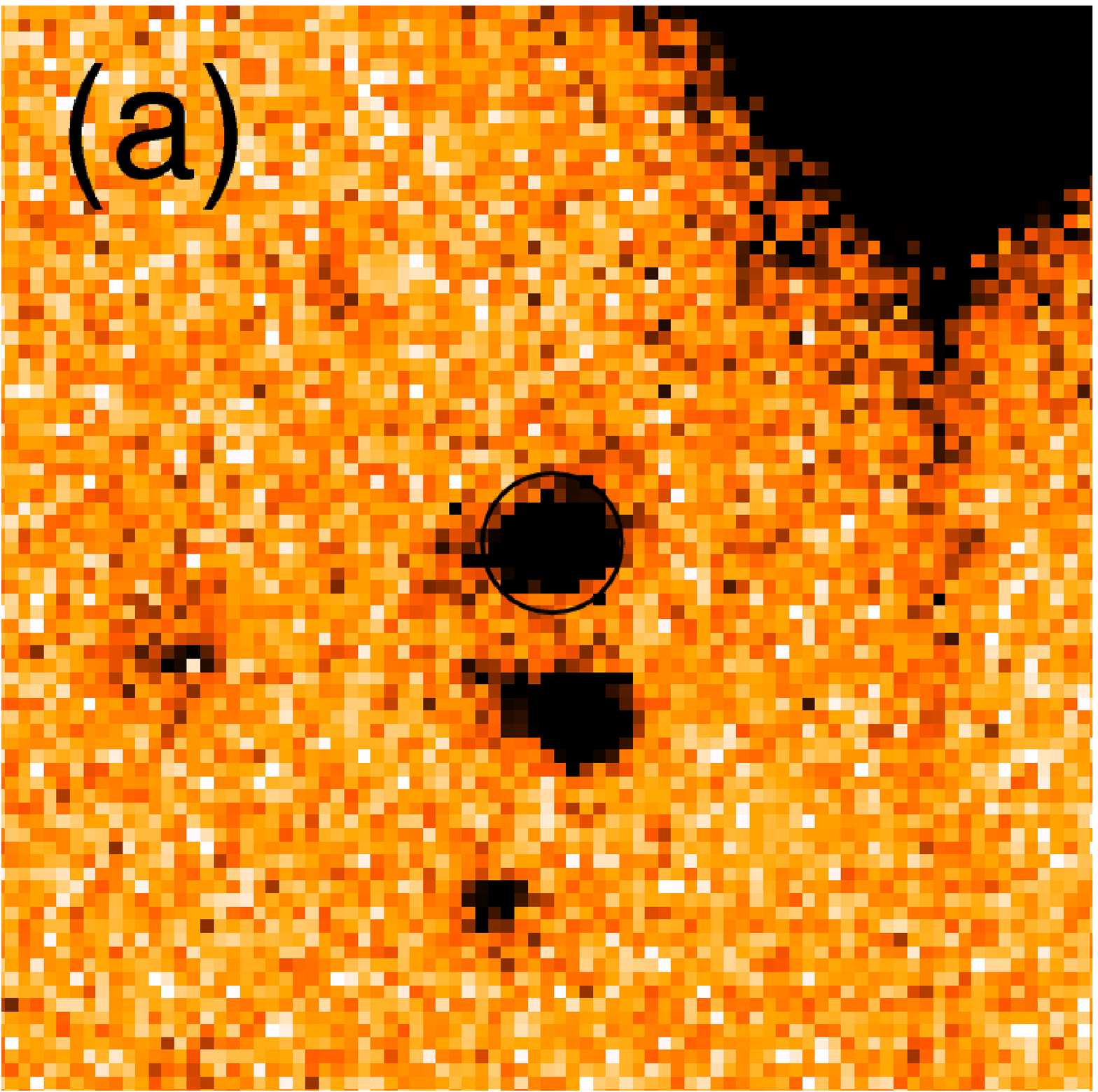}}
\scalebox{0.25}[0.25]{\includegraphics{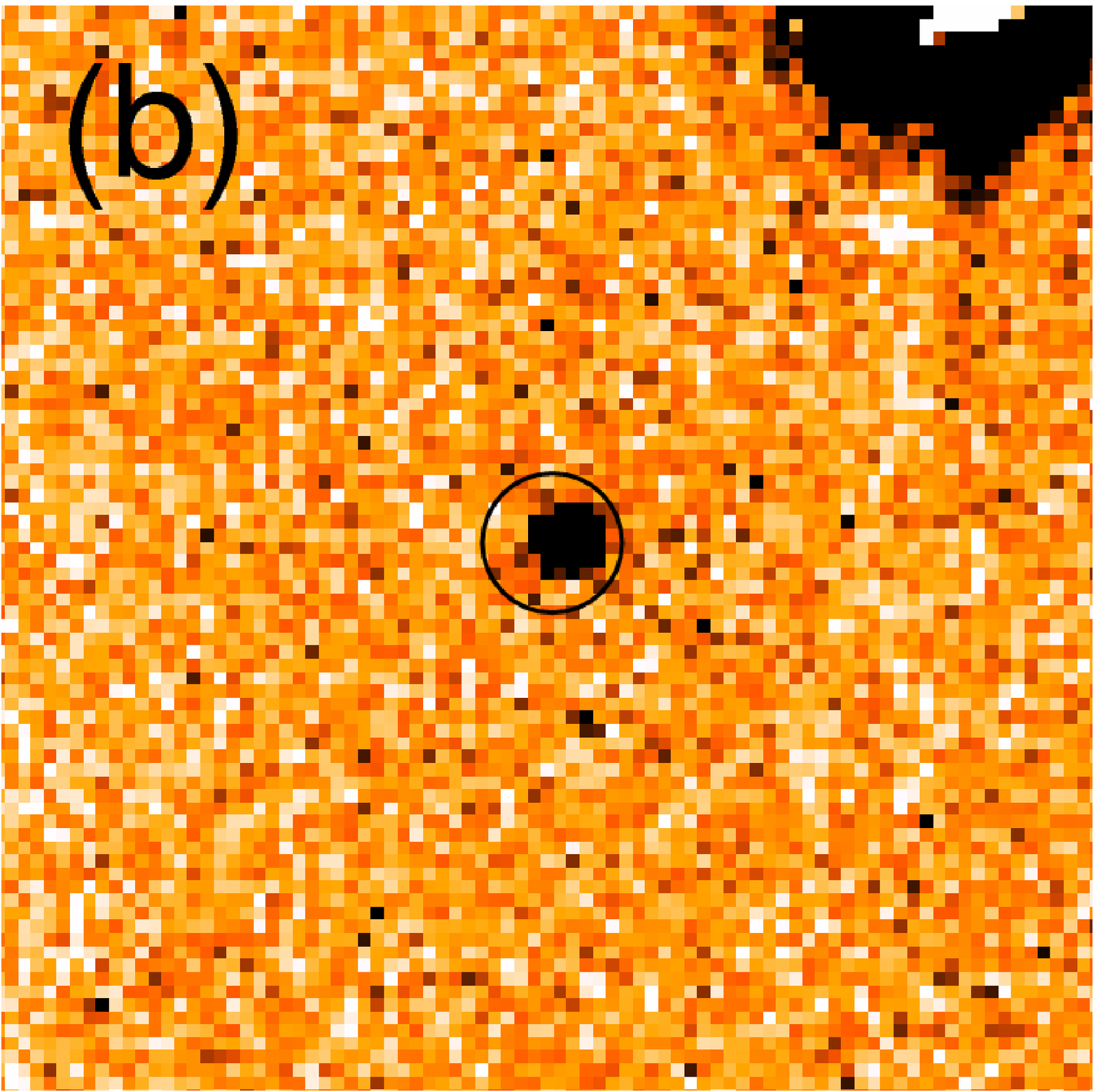}} 
\scalebox{0.25}[0.25]{\includegraphics{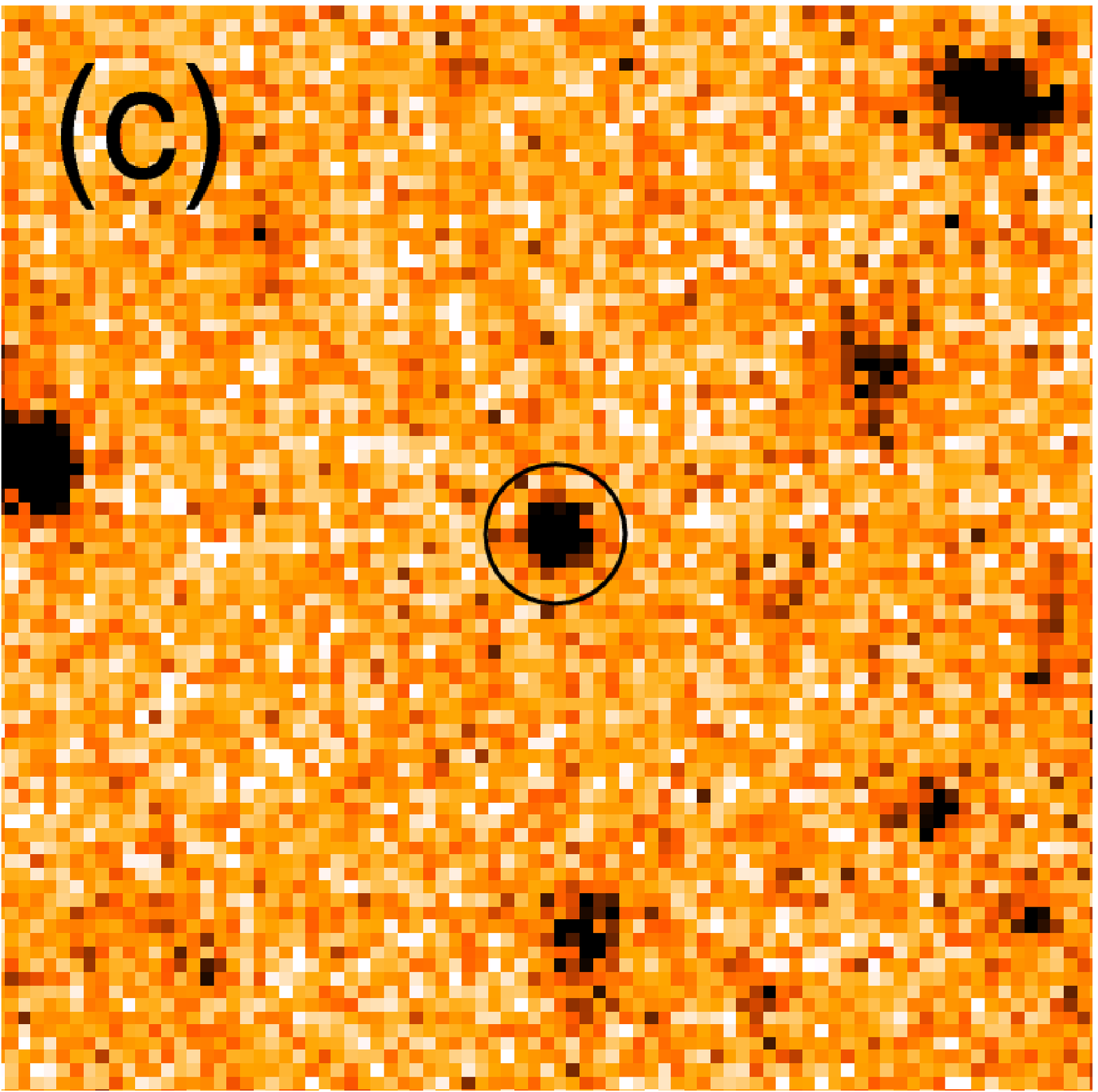}} 
\scalebox{0.25}[0.25]{\includegraphics{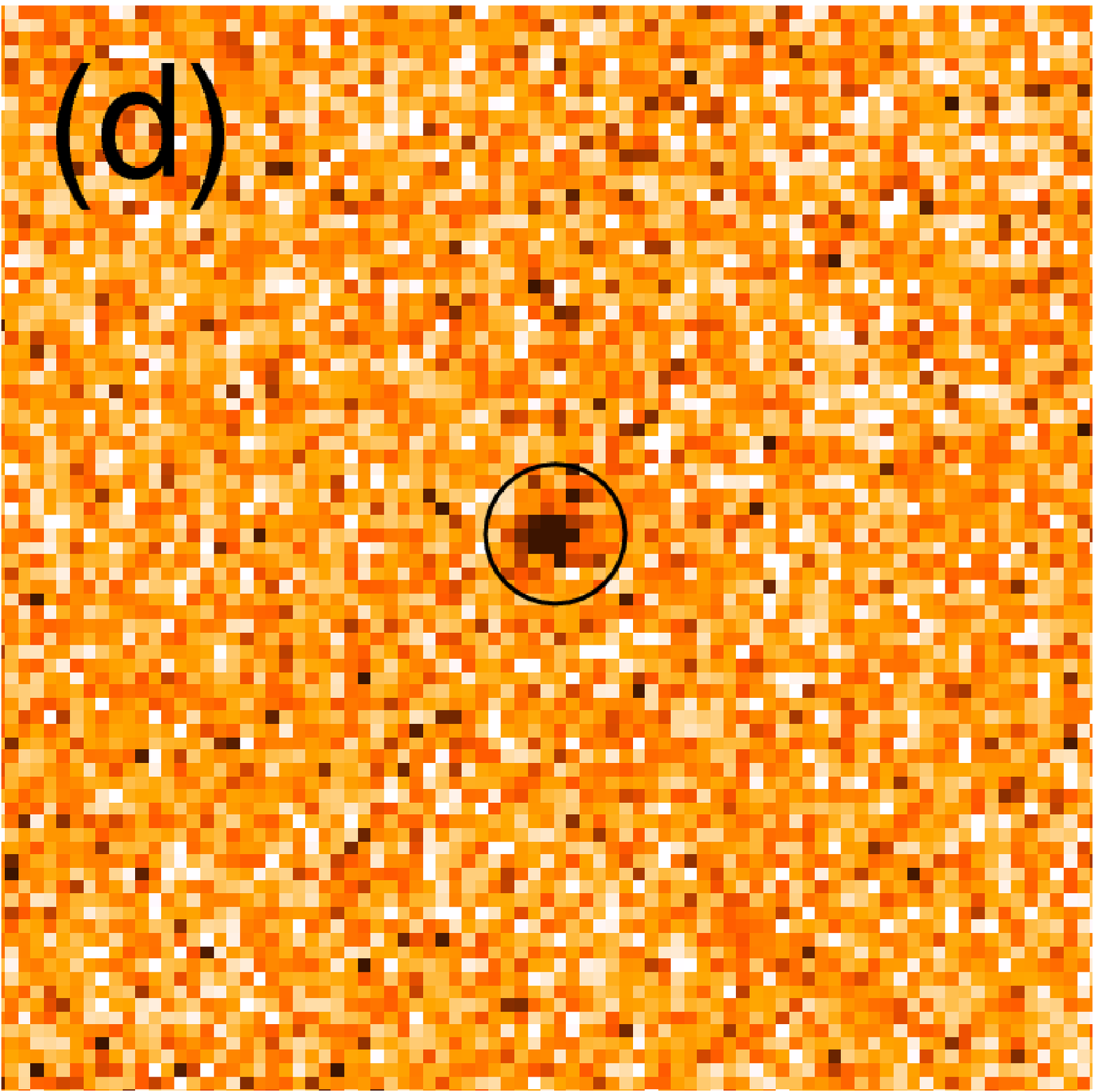}} 

\caption {Thumbnail {\it r'}-band images of the supernovae and their
host galaxies.  All images are centred on the host galaxy FUV centroid
and are 15 arcsec on a side oriented with North upward and East to the
left. {\bf (a):} The SN2213-1745 host galaxy in non-detection season
2004.  The nearby companion galaxy to the South has the same redshift
within the uncertainties and a projected separation of 21 kpc,
physical (\S E).  The proximity of this galaxy and the faint extended
features of the two galaxies suggest that they are interacting.  The
circle is placed on the host galaxy FUV centroid and has a radius of 1
arcsec.  {\bf (b):} Image of SN2213-1745 in the first detection season
2005 after subtraction of the reference image.  The high-quality
CFHTLS yearly stacked images enable clean subtraction images; only the
saturated bright, m$_{\mbox{\scriptsize r'}}$ = 16.7, star in the
upper right hand corner cannot be well-subtracted.  The circle helps
to visualise the potential small offset ($\sim$1 kpc, physical) of the
supernova with respect to the host galaxy centroid.  {\bf (c) \& (d):}
Thumbnail images similar to (a) \& (b) but for the SN1000+0216 host
galaxy in non-detection season 2005/2006 (c) and detection season
2007/2008 (d).}

\label{thumbs}
\end{center} 
\end{figure}

SN2213-1745 was detected in a m$_{\mbox{\scriptsize g'}}$ = 23.4 LBG
in the CFHTLS Deep field D4 and SN1000+0216 was detected in a
m$_{\mbox{\scriptsize i'}}$ = 24.8 LBG in Deep field D2.
Figure~\ref{thumbs} presents small sections of the square-degree
yearly stacked images centred on the host galaxies during
non-detection years and images during detection years with reference
images subtracted to reveal the supernovae.  LBGs at \zz\ - 4 are
essentially point sources in ground-based images (with some showing
extended emission at {\it z} $\lesssim$ 2) with average half-light
radii of $\sim$2 - 3 kpc as determined from space-based
imaging.\cite{gardner00,ferguson04} Our detection technique is biased
towards finding supernovae within the flux extent of the host galaxies
determined from the deep four-year stacked images.  The physical
scales at \zz\ to \zzzz\ range from $\sim$8.6 to $\sim$6.8 kpc
arcsec$^{\mbox{\scriptsize -1}}$ and the plate scale of CFHTLS MegaCam
imager is 0.187 arcsec pixel$^{\mbox{\scriptsize -1}}$, corresponding
to $\sim$1.3-1.6 kpc pixel$^{\mbox{\scriptsize -1}}$ over the
redshifts of interest here.  We note that the separations in our full
sample to date of 12 {\it z} $>$ 2 FUV-luminous supernovae range from
$\sim$0 - 3 kpc, physical, and are consistent with this expectation.

\noindent The superb CFHTLS image astrometry enables relative centroid
accuracies of $\sim$0.04 pixels {\it (\small
~http://www3.cadc-ccda.hia-iha.nrc-cnrc.gc.ca/community/CFHTLS-SG/docs/quality/astrometry.html~\normalsize)}\\
and we find that the relative supernova-host galaxy centroid
measurements season to season and filter to filter are accurate to
$\lesssim$0.5 pixels$^{\mbox{\scriptsize 8}}$ over the full square
degree images.  The relatively central location of the host LBGs in
each field, where astrometric solutions are typically more accurate,
produces centroid accuracies of $\sim$0.3 pixels.  Centroid
separations are measured for each filter between the host centroid in
the two non-detection years and the supernova centroid in the two
detection years.  The measurements produce consistent centroid
separations, with an average {\it g'r'i'} separation of 0.52 $\pm$0.34
kpc, physical, for SN2213-1745.  The small separation is consistent
with zero considering the uncertainties, however, the centroid offsets
appear in the same direction (West) for all 12 measurements with all
errors being consistent.  The average {\it g'r'i'} separation for SN
1000-0216 is 0.62 $\pm$0.57 kpc and consistent with zero.  Here, the
offsets appear in the same direction for the three filters again, but
this time to the South East.  However, the {\it g'}-band detection is
very weak owing to the effects of the \lya\ forest and the Lyman
limit.

\noindent Taking the separations measured from the images at face
value, the two SLSNe appear to reside at, or near, their apparent host
centres.  The two SLSN host galaxies are more luminous than average
LBGs in our sample at the respective redshifts and SLSNe residing near
the host nuclei of luminous host galaxies is a topic of discussion for
low redshift events$^{\mbox{\scriptsize 2,}}$\cite{drake11}.  However,
we note two important points.  Firstly, the image separations are
projected separations in the plane of the sky, thus the true
three-dimensional separations between the LBGs and supernovae are
equivalent to, or larger than, the projected separations.  Secondly,
the images probe the restframe FUV that traces high star formation
regions in the relatively young, high redshift galaxies.  Such regions
are typically clumpy and may, or may not, occur at the host centre or
appear symmetric in intensity about the host centre.  Although the
supernova centroids are reliable, the host galaxy centroids measured
from the images are less so.  As a result, the image separations
presented here should be considered indicative as opposed to
definitive.


\section{Rates}
\label{rates} 

As discussed in the main Letter, SLSNe are classified into three
types, SLSN-I that show no hydrogen, SLSN-II that are hydrogen rich,
and SLSN-R that are powered by $^{\mbox{\scriptsize 56}}$Ni decay and
are likely pair-instability supernovae.  The SLSN rate at low to
intermediate redshift is estimated to be $\sim$10$^{\mbox{\scriptsize
-8}}$ {\it h}$_{\mbox{\scriptsize 71}}^{\mbox{\scriptsize 3}}$
Mpc$^{\mbox{\scriptsize -3}}$ yr$^{\mbox{\scriptsize -1}}$, in which
$\sim$0.2 are SLSN-R candidates$^{\mbox{\scriptsize 1,2}}$.  This
rate is to be compared to 10$^{\mbox{\scriptsize -4}}$ {\it
h}$_{\mbox{\scriptsize 71}}^{\mbox{\scriptsize 3}}$
Mpc$^{\mbox{\scriptsize -3}}$ yr$^{\mbox{\scriptsize -1}}$ for the
full set of core-collapse supernovae\cite{dahlen99,botticella08}
($\gtrsim$8 {\it M}$_\odot$ progenitors).

\noindent We search for SLSNe in LBGs over four redshift paths, \zz,
\zzz, \zzzz\, and {\it z} $\sim$ 5, using standard colour-selection
techniques$^{\mbox{\scriptsize 22-24}}$ tailored to the CFHTLS Megacam
filters.  We map the evolution of 10 starforming and passive galaxy
templates from {\it z} = 0 - 6 in various colour-colour planes to
efficiently ($\gtrsim$90\%) select LBGs at the desired redshifts that
are confirmed from $>$100 LBG spectra acquired specifically for this
programme.  Relevant to the two discoveries presented here, the
colour-selection probes $\delta${\it z} $\sim$ 0.6 at \zz\ and \zzzz.
Computing the effective area of the images and correcting the redshift
paths by the \zz\ and \zzzz\ photometric selection functions results
in volume of $\sim$6 $\times$ 10$^{\mbox{\scriptsize 6}}$ {\it
h}$_{\mbox{\scriptsize 71}}^{\mbox{\scriptsize -3}}$
Mpc$^{\mbox{\scriptsize 3}}$ at both redshifts, adopting an {\it
H}$_{\mbox{\scriptsize 0}}$ = 71, $\Omega_\Lambda$ = 0.73, $\Omega_M$
= 0.27 cosmology.

\noindent The high redshift volumetric rate is estimated by simulating
the rapid evolution of SLSN-I and SLSN-II luminosity profiles and
slower evolving SLSN-II with interaction and SLSN-R profiles at the
two redshifts.  We convolve the supernovae with the detectability
window functions (that extend beyond the formal four observing
seasons) and use the requirement that supernovae need only one season
detection.  Using the specifics of the survey and the four fields, the
one \zz\ detection implies a rough SLSN volumetric rate of $\sim$4
$\times$ 10$^{\mbox{\scriptsize -7}}$ {\it h}$_{\mbox{\scriptsize
71}}^{\mbox{\scriptsize 3}}$ Mpc$^{\mbox{\scriptsize -3}}$
yr$^{\mbox{\scriptsize -1}}$.  The single \zzzz\ detection, along with
the competing effects of shorter time, restframe, sampled by the
CFHTLS photometry and shorter gaps in coverage produces a similar rate
of $\sim$4 $\times$ 10$^{\mbox{\scriptsize -7}}$ {\it
h}$_{\mbox{\scriptsize 71}}^{\mbox{\scriptsize 3}}$
Mpc$^{\mbox{\scriptsize -3}}$ yr$^{\mbox{\scriptsize -1}}$.  We note
that because our supernovae are detected in the FUV, the estimated
high redshift rate reflects a lower limit.  FUV radiation is highly
susceptible to extinction from metal-line absorption and dust in the
local environment of the supernovae, the interstellar medium of the
host galaxies, and the circum-galactic and intergalactic medium.
Thus, there may be a significant fraction of SLSNe with escaping FUV
flux that falls below our detection threshold.

\noindent The rate of core-collapse supernovae closely follows the
star formation rate over cosmic time because of the short lifespans of
massive stars.  The core-collapse supernova rate is expected to
increase by a factor of $\sim$5 from {\it z} $\sim$ 0.3 -
2,\cite{dahlen99,botticella08} following the evolution in the cosmic
star formation rate, and then remain relatively constant or experience
a decline to {\it z} $\sim$ 4.\cite{madau98,hopkins06} The rate of
high-redshift SLSNe based on our two FUV detections, uncorrected for
FUV extinction, is an order of magnitude higher than the rate of SLSN
at low redshift, after correcting for the expected $\gtrsim$5 increase
as a result of increases star formation.  Our rough high redshift SLSN
rate provides the first direct evidence that the number of massive
stars in the early Universe may be higher than that observed at low
redshift.


\section{Light curves}
\label{lcs}

\noindent To generate the supernova light curves for each filter, we
use the technique of point-spread function (PSF) photometry on image
subtractions.  We first construct a reference image from data where no
supernova light is expected (2003 for SN2213-1745 and 2004 for
SN1000+0216). The reference image is then subtracted from each image
where the supernova light is present using a PSF matching technique,
removing the host galaxy light and leaving only the supernova light in
the difference image. The supernova photometry is then measured with a
PSF-fitting technique that uses the PSF measured from nearby field
stars.\cite{guy10} The photometric data are calibrated using a set of
tertiary standard stars.\cite{regnault09} The photometry for SN
2213-1745 and SN1000+0216 is listed in Tables~\ref{sn224mags}
and~\ref{sn279mags}.

\subsection{Rise times} 
\label{rise}

\begin{figure}[h!]
\begin{center}
\scalebox{0.35}[0.35]{\rotatebox{90}{\includegraphics{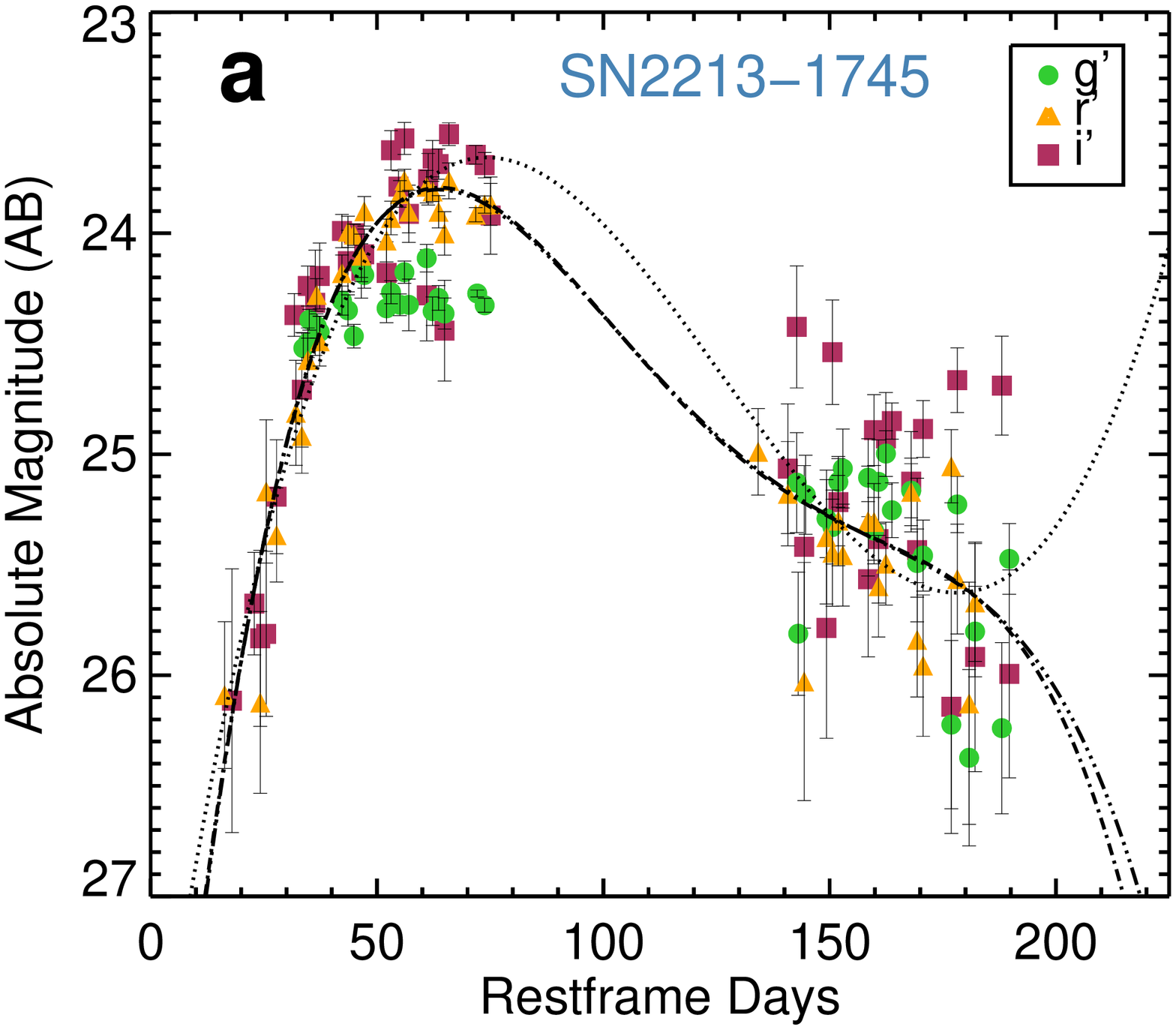}}}
\scalebox{0.35}[0.35]{\rotatebox{90}{\includegraphics{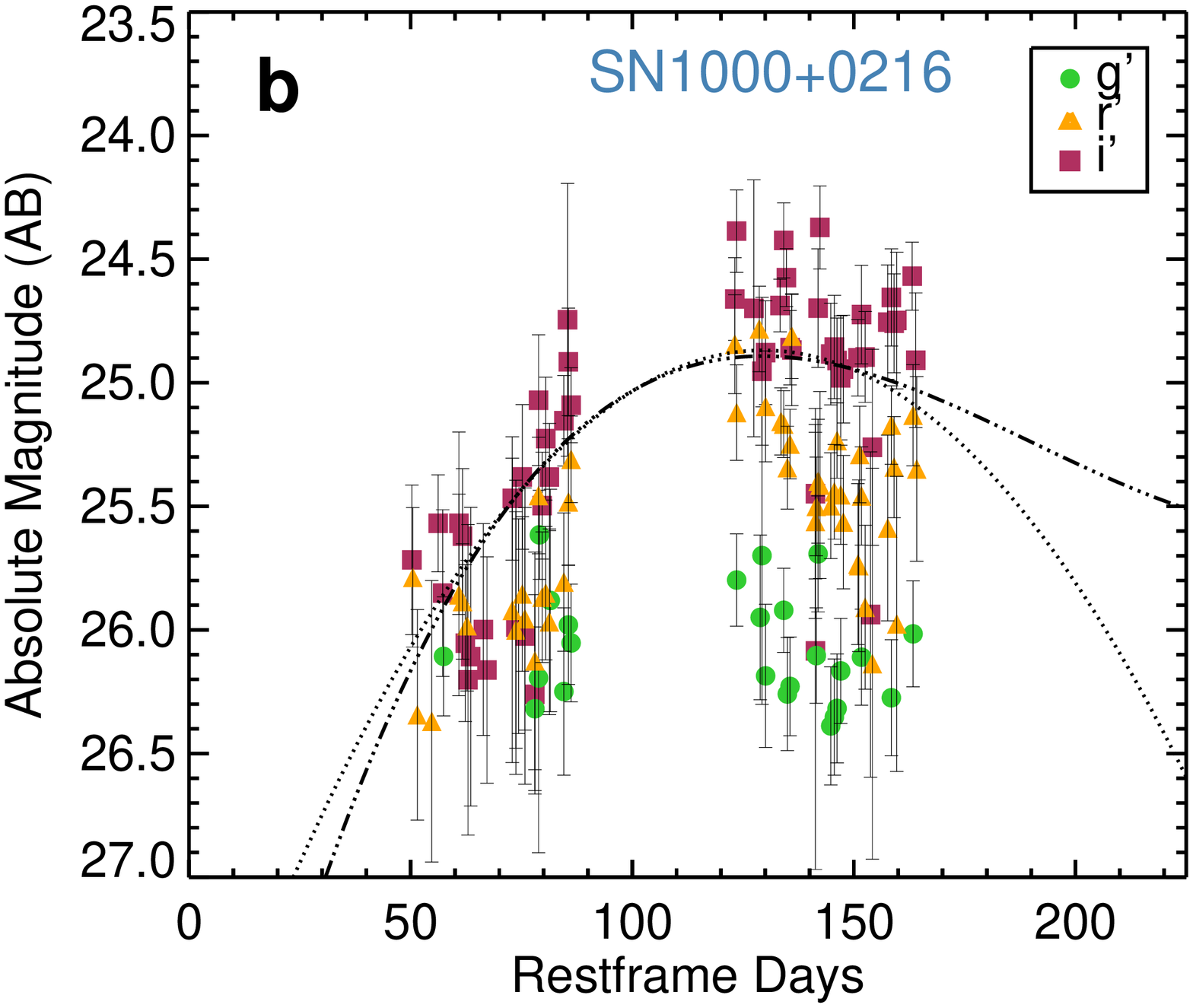}}}

\caption {Polynomial fits to the nightly stacked image photometry for
the two high redshift supernovae.  The gap in coverage between
observing seasons alters the shape of the polynomial and makes the
determination of the time of peak magnitude less reliable,
particularly in the case of SN 2213-1745 which exhibits a larger
change in magnitude over a larger number of restframe days.  {\bf
(a):} The photometry of SN2213-1745 is shown fit with third (dotted
curve), fourth (dot-dot-dash curve), and fifth (dot-dash curve) order
polynomials.  {\bf (b):} Similar to panel (a), but for SN1000+0216.
Because of the effects of the \lya\ forest and the Lyman limit on the
{\it g'} and {\it r'} filters, only the {\it i'}-band that probes the
supernova continuum is fit.  The results from second (dotted curve)
and third (dot-dash curve) order polynomial fits are shown.  }

\label{poly}
\end{center} 
\end{figure}

The photometric data points and the limits imposed by the
non-detections in previous seasons indicate $\sim$50 to $>$150 day
rise time from outburst to peak magnitude for the two supernovae.  As
an exercise to help narrow the rise time range and possibly shed light
on the nature of the supernovae, we first follow a conventional
approach and fit low order polynomials to the slowly evolving
photometric data. The polynomial fit approach estimates the date of
outburst and rise time to peak without any assumptions of the
progenitor however, for the CFHTLS seasonal data, has difficulty
fitting the evolution over the $\sim$6 month periods, observed frame,
in which the supernovae were not observed.  The peak magnitudes are
not well constrained within the coverage gaps and it is unclear
whether the peak for SN1000+0216 falls within the gap or occurs beyond
the extent of the data.  As a result, the polynomial fit rise times
are rough estimates and defined here as the time for the supernovae to
rise 5 magnitudes to peak as determined by the functional fit.
Figure~\ref{poly} displays the polynomial fits for both supernovae.
In addition, because the high redshift supernova data follow the
behaviour of low-redshift SLSNe-R (see main Letter), we compare their
light curve evolution to pair-instability supernova models.

\noindent {\bf Polynomial fits} The polynomial fits to the SN2213-1745
{\it g', r',} and {\it i'} data separately produce very similar
results, but the lack of data constraints during rise for the {\it g'}
data yields a significantly slower rise time ($>$200 days) for a third
order polynomial.  Because the {\it r' and i'} data follow a similar
evolution over the date ranges probed by each filter, the values
reported here are those fit to the combined data from the {\it r' and
i'} filters.  The SN2213-1745 {\it r'i'} data are fit with third,
fourth, and fifth order polynomials and result in rise times of 81,
62, and 64 days, respectively.  We note that these rise times are a
lower limit to the true value when considering that the observational
gap of photometric coverage occurs after these estimated dates of peak
magnitude.  We adopt an intermediate polynomial fit rise time of 70
days for this supernova.

\noindent Second and third order polynomial fits to the {\it i'}
filter data of SN1000+0216 result in rise times of 146 and 129 days,
respectively (higher polynomials were unconstrained at early times).
Second and third order fits to the {\it g'} filter appear identical,
with each producing rise times of 134 days.  A second order fit to the
{\it r'} filter produces a rise time of 138 days with the third order
fit unconstrained at early times.  The {\it g'} and {\it r'} fits are
not shown in Figure~\ref{poly} for clarity.  If we assume that the
true peak magnitude occurs between detection seasons, the
non-detection in the previous season places an upper limit on
detection to $\sim$130 days.  Otherwise, the rise time is $>$150 days
if the peak occurs beyond the data.  Given that the {\it g'} and {\it
r'} filters probe the \lya\ forest and are detected at lower
significance, we focus on the fits to the {\it i'} filter and adopt a
polynomial rise time of 129 days.

\begin{figure}[h!]
\begin{center}
\scalebox{0.35}[0.35]{\rotatebox{90}{\includegraphics{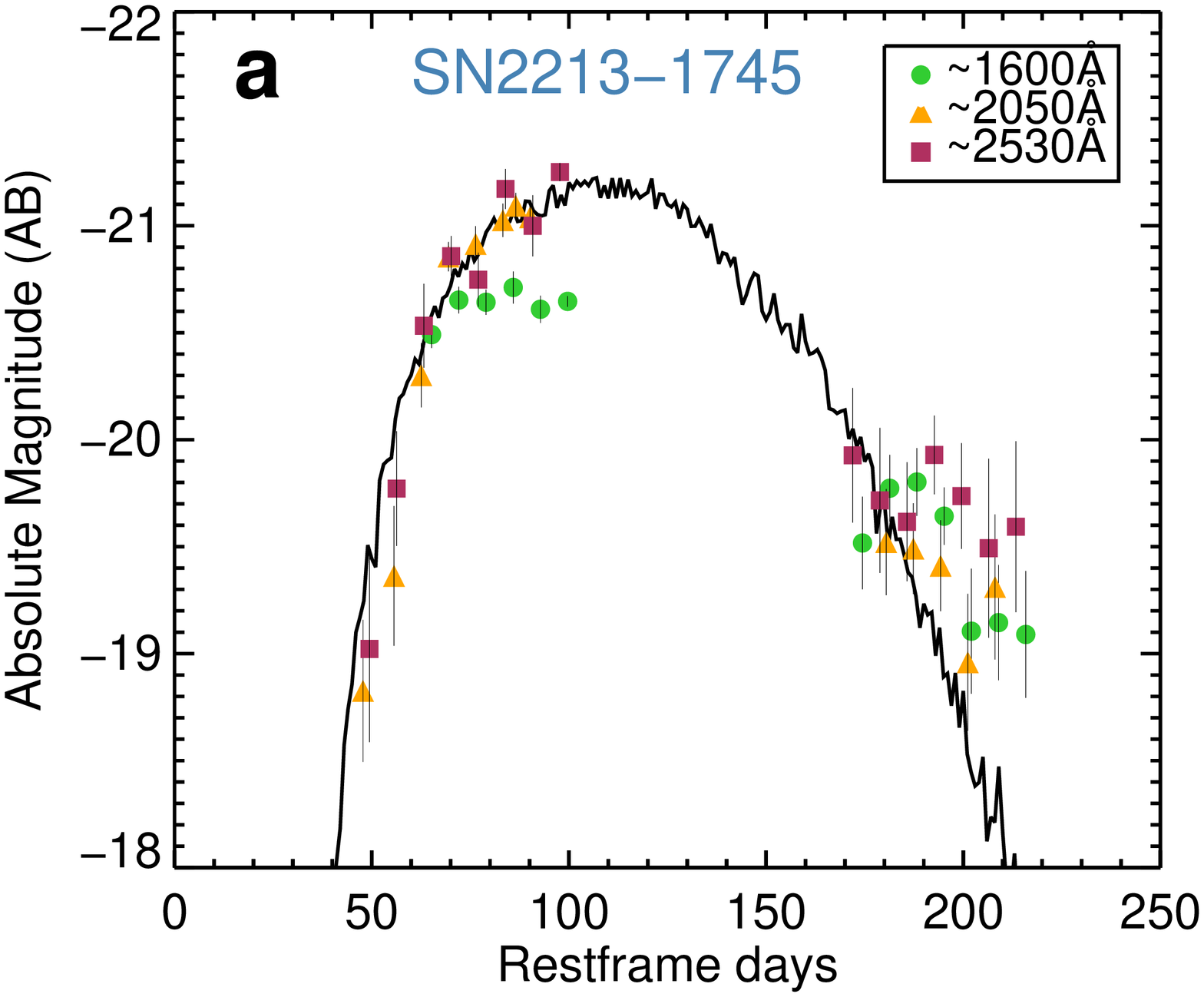}}}
\scalebox{0.35}[0.35]{\rotatebox{90}{\includegraphics{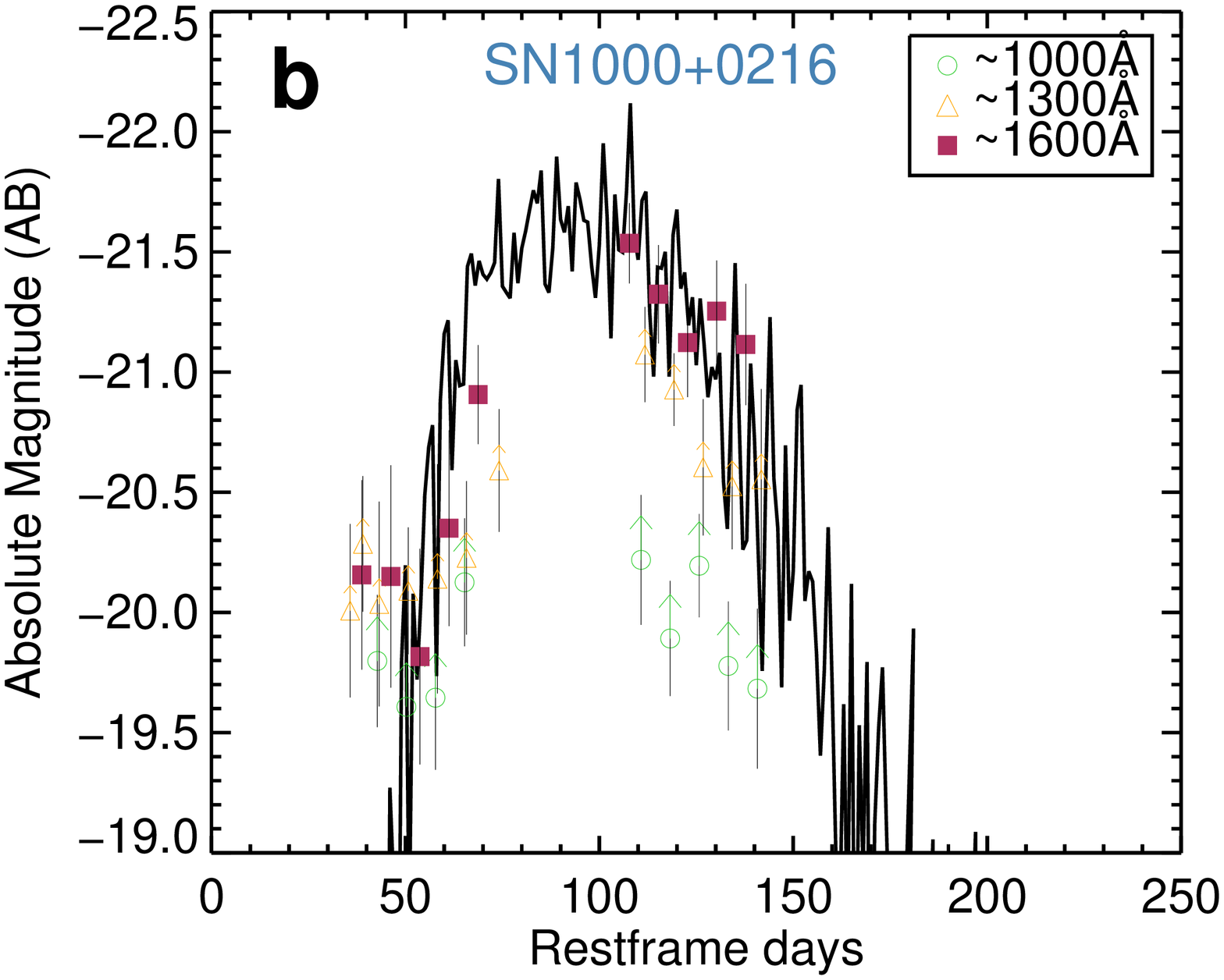}}}

\caption {Comparison of the data to the time evolution of a
pair-instability supernova model.  {\bf (a)} The rise and decay of
SN2213-1745 binned in 7-day intervals.  The overlaid curve is the
far-ultraviolet evolution of a 130 {\it M}$_\odot$ bare helium core
pair-instability supernova model averaged over the restframe continuum
near 2000\AA~as sampled by the observed {\it r'} bandpass and offset
to match the luminosity of the data {\it (see text)}.  {\bf (b)}
Similar to panel (a) but for SN1000+0216.  Because the {\it g'} and
{\it r'} filters probe the \lya\ forest and are subject to significant
absorption by gas from lower redshift systems in the line of sight
(the true flux is higher as denoted by the upward arrows), the model
is averaged over the restframe continuum near 1600\AA~as sampled by
the observed {\it i'} bandpass.  The supernovae appear to follow the
far-ultraviolet evolution of the model well, however the higher
luminosity of the data may point to either model limitations or a
different (or additional) luminosity mechanism. }

\label{model}
\end{center} 
\end{figure}

\noindent {\bf Pair-instability supernova models} Stars with masses
between 140 - 260 {\it M}$_\odot$ are theorised to end their lives as
pair-instability supernovae (PISNe).$^{\mbox{\scriptsize 4-6}}$ The
high temperatures and relatively low densities that develop in the
cores of such massive stars promote electron-positron pair production.
The rapid conversion of pressure-supporting radiation into rest-mass
in such a core leads to a hydrostatic instability that can trigger a
run-away nuclear explosion that completely obliterates the star.
PISNe produce peak energies of $\gtrsim$10$^{\mbox{\scriptsize 44}}$
erg and exhibit a slow light curve evolution consistent with the
radiative decay of $^{\mbox{\scriptsize 56}}$Ni.  Because this
behaviour is similar to that of our high redshift SLSNe, we compare
their FUV evolution to that of radiative hydrodynamics PISN
simulations.$^{\mbox{\scriptsize 19}}$

\noindent We find that 150 - 250 {\it M}$_\odot$ low-metallicity
stellar models that retain their hydrogen envelopes (blue and red
supergiant models) evolve too slowly to peak luminosity ($\gtrsim$200
days) and/or have too low FUV energies to provide a reasonable fit to
the data.  We note that we cannot rule out the more luminous supergiant
models for SN1000+0216 if the peak occurs beyond the extent of the
data.  We find good fits to a 130 {\it M}$_{\odot}$ bare helium core
model that follows the evolution of progenitor star with an initial
mass of $\sim$250 {\it M}$_\odot$ with its outer envelope, consisting
largely of hydrogen, removed from binary mass exchange or stellar
evolutionary processes.

\noindent The evolution of SN2213-1745 follows the form of the bare
helium core model well, as seen in Figure~\ref{model}, and does so for
the model integrated over each of the restframe wavelength ranges that
corresponding to the the {\it g',r'}, and {\it i'} filters; 1600\AA\
$\pm$240 ({\it g'}), 2060\AA\ $\pm$200 ({\it r'}), and 2550\AA\
$\pm$250 ({\it i'}).  The model suggests a rise time of $\sim$105
days, however, the supernova is more luminous than expected by a
factor of 2.8, 1.9, and 0.7 for the {\it g',r'}, and {\it i'} filters,
respectively.  Although the model is the state-of-the-art for PISN
simulations, it remains less well constrained in the FUV.  The higher
luminosity and bluer colour of SN2213-1745 may be a result of higher
photospheric explosion temperatures and/or simulation limitations.
The model resolution for the shorter wavelengths probed by SN1000+0216
data is more limited.  We integrate the model over the longer
wavelengths probed by the {\it i'} filter, 1600\AA\ $\pm$150, that
samples the supernova continuum and avoids the effects from the \lya\
forest and Lyman limit.  We find that SN1000+0216 appears to follow
the evolution of the model reasonably well, peaking at $\sim$95 days
after outburst, but it is $\sim$2 mag more luminous.

\noindent From the PISN model, a low-order polynomial fit to the data,
and the constraints set by the previous season non-detections, we
estimate a 75 -- 125 day rise time from outburst to peak magnitude for
SN2213-1745 and 95 -- 130 days for SN1000+0216.


\section{Nature of the High-Redshift Super-luminous
Supernovae}
\label{nature}

\noindent In an attempt to understand the nature of the high redshift
SLSNe, it is important to keep in mind that the light curves of the
supernovae do not match the quicker fade rates of SLSN-I or SLSN-II at
low and intermediate redshift (cf. Figure 2, main Letter).  Moreover,
our SLSNe are observed in the FUV where the fade rates are expected to
evolve quicker than the near-UV and optical.  Instead, the evolution
of the two high redshift supernovae provide a good match to that of
SLSN-R and the 0.0098 mag day$^{\mbox{\scriptsize -1}}$decay rate of
$^{\mbox{\scriptsize 56}}$Co (daughter product of $^{\mbox{\scriptsize
56}}$Ni) out to the extent in which light curves are sampled.  We find
that the observed high FUV luminosities may push the supernovae beyond
the energies possible for the PISN interpretation of SLSN-R as
estimated by the bare helium core model discussed in the previous
section, in particular for SN1000+0216.  A possible alternative and/or
additional source for late-time emission is the interaction of the
supernovae with previously expelled circumstellar material
(\S\ref{SNem}).

\noindent One plausible interpretation for the high redshift SLSNe is
that they are examples of pulsational pair supernovae.  Stars in the
mass range $\sim$80 - 150 {\it M}$_\odot$, depending on rotation rate,
metallicity, and other parameters, may experience pulsational pair
instability and fail to completely explode on their first
try.\cite{woosley07} During the evolution of a star this massive, the
core becomes pair-instable and the explosion ejects many solar masses
of outer envelope material but the energy is insufficient to unbind
the star.  The core stabilises again, contracts, and grows hotter.  A
subsequent pair-instability pulsation drives off additional material
at higher velocities than the first ejection that collides with the
originally expelled material, leading to an extremely luminous
event$^{\mbox{\scriptsize 19}}$.  The expected higher rate of
pulsational pair-instability supernovae as compared to PISNe is more
consistent with our rough rate estimate, given the acceptable mass
range of each population, however, such events do not generate the
large amounts of $^{\mbox{\scriptsize 56}}$Ni needed to power the
observed late-time luminosity.  Similar to type IIn supernovae,
collision of shells of material could, in principle, provide an
additional late-time luminosity contribution and generate
shock-induced emission that persists to late times.  Evidence for
long-lived circumstellar interaction-induced emission features can be
detected in deep, late-time FUV spectroscopy in some cases and is
discussed in \S\ref{SNem}.


\section{Host galaxies}\label{hosts}

\subsection{SN2213-1745}
\label{224556}

The host galaxy to SN 2213-1754 is a luminous, M$_{\mbox{\scriptsize
FUV}}$ = -21.4, \zz\ LBG.  The star formation rate (SFR) of the host
galaxy based on the restframe FUV (AB)$^{\mbox{\scriptsize 10,11}}$
magnitude derived from the observed {\it g'} band (restframe
$\sim$1600\AA) magnitude, is estimated to be $\sim$20 {\it M}$_\odot$
yr$^{\mbox{\scriptsize -1}}$ when using the simple relationship, SFR =
6 $\times$ 10$^{\mbox{\scriptsize -(8+0.4 M(FUV))}}$ {\it M}$_\odot$
yr$^{\mbox{\scriptsize -1}}$, that ignores the effect of
dust.\cite{kennicutt98,madau98} The host was densely monitored
photometrically for four observing seasons and generated a single
supernova-like outburst.  The host spectrum shows features that are
representative of typical LBGs, including a strong \lya\ absorption
feature similar to that seen in $\sim$25\% of the LBG population, and
no features typically attributed to AGN.  An LBG composite template
gives a very good fit to the spectrum as seen in
Figure~\ref{hostspec}.

\noindent The object located 2.45 arcsec (20.7 kpc at {\it z} $\sim$
2) to the South of the SN2213-1745 host galaxy (see
Figure~\ref{thumbs}) has a similar apparent magnitude, m$_{r'}$ =
23.8, and is also a colour-selected \zz\ LBG.  The spectroscopic
observations for the SN2213-1745 were executed such that the longslit
was oriented to simultaneously obtain the flux from the nearby object.
The observations were done without loss of wavelength-dependent flux,
which could occur if the orientation of the two objects is not optimal
with the parallactic angle, because the atmospheric dispersion
corrector on LRIS was deployed.  The spectroscopy confirms the object
as an LBG with redshift {\it z} = 2.0455 $\pm$0.0008.  Inspection of
the flux contours of the CFHTLS images reveals diffuse flux extending
to the East of the two LBGs similar to the appearance of tidal tails,
providing evidence that the two galaxies are interacting.  Galaxy
interactions induce the collapse of cold gas and trigger star
formation in galaxies and associated gas
clouds.\cite{larson78,barton00,barton03} The star forming episodes
that result from interactions increase the chance of observing the
deaths of short-lived very massive stars.

\subsection{SN1000+0216}
\label{279200}

The host to SN1000+0216 is a luminous, M$_{\mbox{\scriptsize FUV}}$ =
-21.2, \zzzz\ LBG.  Using the SFR relationship above, the similar FUV
luminosity to SN2213-1745 (derived from the observed {\it i'}-band
magnitude, restframe $\sim$1600\AA\ at \zzzz, similarly produces a FUV
SFR of $\sim$20 {\it M}$_\odot$ yr$^{\mbox{\scriptsize -1}}$.  The
host exhibited a single supernova-like outburst over the four years
monitored and we see narrow \lya\ in emission that is common to
$\gtrsim$50\% of the LBG population, and no associated emission
attributable to AGN activity.  Finally, there is no clear evidence for
galaxy interaction as there is for the host of SN2213-1745.  We note
that the depth and resolution of the data are not sensitive to late
stages of interaction, interactions with minor companions
(approximately 5:1 or less mass) which are more common, or interaction
with companions that are not currently undergoing starbursts (e.g., on
their first pass) and thus do not produce strong FUV flux as is
witnessed for low-redshift LBG
analogues.\cite{overzier08,overzier10,mannucci09}

\subsection{SLSNe-R and Population III star hosts} \label{popIII}

Theory states that massive stars form more easily in regions of
low-metallicity whereas Population III stars form in regions of
essentially zero metallicity.  At low redshift, the environments that
best suit the necessary conditions for massive star formation may be
low-luminosity, low-metallicity star forming dwarf galaxies.  Indeed,
low-redshift SLSNe, including SLSNe-R, are found in star forming
galaxies with luminosities -16 $<$ M $<$ -21, with two luminous
exceptions, SCP 06F6 and SN 2006gy.\cite{neill11} In addition, the
hosts have relatively low metallicities (where measured) and high
specific star formation rates (star formation rate per unit galaxy
stellar mass).  The low-redshift SLSN-R SN 2007bi was detected in a
faint (M$_{\mbox{\scriptsize B}} =$ -16), star forming host
galaxy$^{\mbox{\scriptsize 13}}$ with estimated low metallicity
($\sim$ 0.3 solar)$^{\mbox{\scriptsize 17}}$.

\noindent Our high-redshift supernova hosts are star forming galaxies
residing at the luminous end of the low-redshift SLSN host galaxy
luminosity distribution.$^{\mbox{\scriptsize 48}}$ We remark that our
survey is not sensitive to host galaxies with M$_{\mbox{\scriptsize
FUV}} \gtrsim$ -19.5, but a search for high redshift `orphan'
supernovae, events whose host galaxy is below the detection threshold
of the survey, is in progress.  Our FUV data alone do not provide
reliable metallicities for our supernova hosts but the typical
metallicities of LBGs measured using infrared (restframe optical)
spectroscopic diagnostics are low, $\sim$0.5 - 0.1
solar.\cite{mannucci09} Finally, the two high redshift supernova hosts
have high specific star formation rates which is a ubiquitous feature
of the LBG population.  Thus, the low metallicities and high specific
star formation rates are consistent with low redshift SLSN-R hosts,
but the relatively high luminosities of the high-redshift host
galaxies would seem at odds with that of SN 2007bi.  However, it is
unlikely that the potential low-redshift preference for faint SLSN-R
hosts similarly holds at high redshift where typical galaxies have
low-metallicities and high star formation rates, thus higher intrinsic
luminosities.

\noindent The many ISM metal absorption lines in our deep spectra
reveal that the globally averaged metallicity of the high redshift
supernova host galaxies is not primordial.  Consequently, in order for
the supernovae to be Population III stars, they would need to have
formed in pockets of pristine gas.  This could occur either within the
galaxies (e.g., the enrichment of metals in the galaxies is clumpy),
in the low metallicity environment of the galaxy haloes, or in
sub-haloe galaxies or line-of-sight gas clouds in the host vicinity.
Finally, we note the possibility for SLSNe to form as a result of a
collision of two stars in young, dense star clusters in the host
galaxies.\cite{pan11} The clumpy star forming nature of LBGs could
provide a environment that is conducive to stellar collisions,
although this type of event is predicted to be $\sim$100 times more
rare than our estimated rate.


\section{Supernova late-time emission}
\label{SNem}

\noindent As discussed above, the interaction of circumstellar
material may be a mechanism to contribute to the luminosity of SLSNe.
One example is SN 2006gy, a SLSN-II at low redshift, that exhibited
circumstellar interaction and may be an example of a pulsational
pair-instability event.\cite{woosley07} Such interactions may produce
FUV emission features that are similar to the long-lived features
observed in local type IIn supernovae.\cite{fransson02,fransson05} Our
survey for FUV-luminous supernovae at high redshift has detected type
IIn-like emission features in several of our confirmed {\it z}
$\gtrsim$ 2 supernova spectra.  In this section, we outline our
approach to search for late-time emission in high redshift supernova
spectra.

\noindent LBGs exhibit a prominent \lya\ feature observed in
absorption, emission, or a combination of both.  Apart, from \lya,
LBGs show no significant emission lines over the wavelengths studied
other than weak He\textsc{ii} 1,640\AA\ and C\textsc{iii}] 1,909\AA\
that appear in some spectra.  Our approach is to investigate any
excess emission features in our LBG hosts, particularly at atomic
transitions seen in low-redshift FUV supernova spectra that exhibit
late-time circumstellar interaction.  Features include \lya\ 1,216\AA,
N\textsc{v} 1,240\AA, O\textsc{i} 1,358\AA, O\textsc{iv} 1,400\AA,
N\textsc{iv}] 1,486\AA, excess C\textsc{iv} 1,550\AA, N\textsc{iii}
1,753\AA, Si\textsc{ii} 1,808\AA, NeIII] 1,815\AA, Si\textsc{iii}]
1,883\AA, and excess CIII] 1,909\AA.  An important consideration is
that the peaks of many circumstellar interaction emission features in
low-redshift spectra are blueshifted by $\sim$1000 - 4000 km
s$^{\mbox{\scriptsize -1}}$ and can appear relatively broad.
Moreover, because the high redshift spectra include the entire flux
from the host, any supernova features are affected by interstellar
medium absorption features of the host that can significantly alter
the final observed profile.  This differs from low redshift where the
host galaxy contamination in supernova spectra is small.  Our
programme includes future spectroscopy of host galaxies having
candidate supernova emission to confirm or refute any potential
detections by the decay, disappearance, or continued presence of the
lines.
 
\begin{figure}[!h] \begin{center}
\scalebox{0.34}[0.34]{\rotatebox{90}{\includegraphics{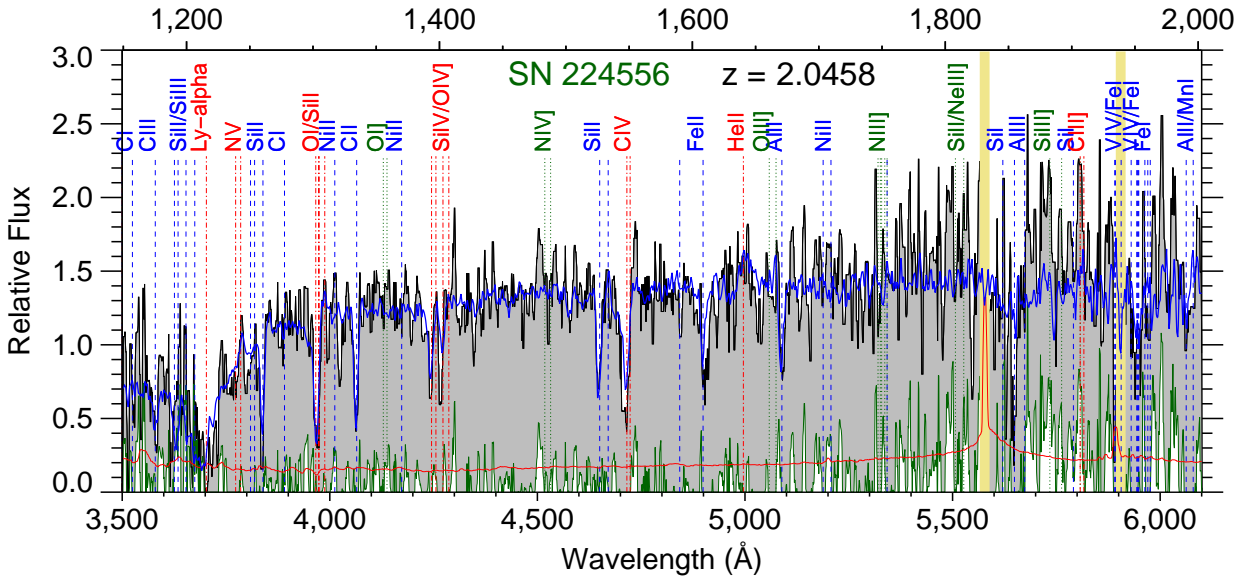}}}

\caption {Late-time host galaxy spectrum similar to Figure 3 in the
main Letter.  SN2213-1745 host galaxy spectrum (black curve with gray
fill) is shown smoothed by 5 pixels.  Labelled vertical (blue) dashed
lines indicate typical atomic transitions seen in absorption in LBGs.
Vertical (green) dotted lines mark late-time transitions observed in
low-redshift supernovae exhibiting circumstellar interaction and are
not observed in LBGs and red (dot-dashed) lines denote transitions
seen in both LBGs (typically in absorption) and supernovae (in
emission).  The host galaxy error array is shown as the thin red curve
and thick vertical (yellow) lines mark the positions of bright night
sky lines.  The difference spectrum, after the subtraction of the
composite spectrum from the data, is shown as the (green) curve near
zero with white fill.  As noted in the text, additional absorption
features in the difference spectrum is expected as a result of lower
redshift systems in the line of sight, but additional emission
features are not.  The lower signal-to-noise ratio and fluxing caveats
(see text) of the existing SN1000+0216 data does not enable a similar
difference spectrum analysis.  }

\label{hostspec}
\end{center}
\end{figure}

\noindent In an effort to isolate potential late-time supernova
emission, we subtract matched composite spectra from the host
galaxies.  The composite spectra are constructed from 200 \zzz\ LBGs
selected based on the observed relationships between \lya\ equivalent
width, FUV continuum, and ISM absorption line
profiles\cite{shapley03}.  As seen in Figure~\ref{hostspec}, the
composite spectrum provides an excellent fit to the relatively high
signal-to-noise ratio continuum and prominent features of the host
galaxy to SN2213-1745.  We note that several absorption features are
seen in the data, that are not seen in the template, that result from
metal line absorption from lower redshift systems in the line of
sight.  These features are also seen in the difference spectrum shown
as the green curve with white fill in the figure.  Although additional
absorption features are expected, additional emission features are
not.

\noindent The SN2213-1745 host galaxy exhibits strong \lya\
absorption, similar to that seen in $\sim$25\% of the LBG population,
and little evidence for hydrogen emission from the supernova that
would classify this event as a SLSN-II.  Although the host LBG might
absorb a significant fraction of the \lya\ emission, the expected
broad nature of the feature as seen in low-redshift type IIn supernova
spectra would predict a leakage of measurable fraction of emission
given the expected clumpy distribution of neutral hydrogen in the
host.  The lack of \lya\ emission provides spectroscopic evidence
against a SLSN-II classification, independent of the mismatch between
the light curve evolution of low redshift SLSN-II and SN2213-1745, and
further supporting a SLSN-R interpretation.

\noindent The SN 1000+0216 host galaxy exhibits \lya\ in emission,
common to $\gtrsim$50\% of the LBG population (see Figure 3, main
Letter).  The lower signal-to-noise ratio and fluxing caveats (\S
\ref{spec}) of the existing data does not allow a similar difference
spectrum analysis to that of SN2213-1745.  As a result, it is
currently unclear whether any excess emission from the supernova is
present in the form of a contribution to the \lya\ feature or other
FUV emission lines typical of circumstellar interaction.  Additional
spectroscopy of SN1000+0216 and future spectroscopy ($>$3 years
post-detection, restframe) of both supernovae will enable a more
comprehensive late-time emission analysis.

\begin{table}[h!]
\centering 
{\footnotesize
\begin{tabular}{lcccccccc}
\hline 
day (MJD) & {\it g'} & error & day (MJD) & {\it r'} & error & 
day (MJD) & {\it i'} & error\\
\hline
53551 & 26.80 & 0.68 & 53529 & 26.81 & 0.91 & 53535 & 26.11 & 0.59 \\
53554 & 27.17 & 0.55 & 53530 & 26.08 & 0.33 & 53550 & 25.67 & 0.23 \\
53583 & 24.52 & 0.06 & 53535 & 29.42 & 2.15 & 53554 & 25.83 & 0.39 \\
53586 & 24.49 & 0.06 & 53554 & 26.12 & 0.41 & 53558 & 25.81 & 0.37 \\
53587 & 24.39 & 0.04 & 53558 & 25.16 & 0.32 & 53565 & 25.19 & 0.25 \\
53592 & 24.42 & 0.05 & 53565 & 25.36 & 0.21 & 53577 & 24.37 & 0.09 \\
53594 & 24.44 & 0.07 & 53578 & 24.81 & 0.23 & 53582 & 24.70 & 0.25 \\
53609 & 24.30 & 0.06 & 53582 & 24.91 & 0.16 & 53586 & 24.23 & 0.09 \\
53613 & 24.34 & 0.07 & 53586 & 24.57 & 0.09 & 53591 & 24.31 & 0.23 \\
53617 & 24.46 & 0.05 & 53592 & 24.27 & 0.07 & 53594 & 24.19 & 0.14 \\
53622 & 24.15 & 0.06 & 53594 & 24.48 & 0.11 & 53609 & 23.99 & 0.07 \\
53624 & 24.18 & 0.05 & 53609 & 24.18 & 0.08 & 53613 & 24.12 & 0.13 \\
53639 & 24.33 & 0.06 & 53613 & 23.99 & 0.07 & 53617 & 23.99 & 0.04 \\
53642 & 24.26 & 0.05 & 53617 & 24.00 & 0.04 & 53622 & 24.17 & 0.12 \\
53648 & 24.32 & 0.04 & 53622 & 24.10 & 0.09 & 53624 & 24.09 & 0.11 \\
53651 & 24.17 & 0.05 & 53624 & 23.90 & 0.06 & 53639 & 24.17 & 0.14 \\
53654 & 24.32 & 0.11 & 53639 & 24.03 & 0.09 & 53642 & 23.62 & 0.08 \\
53666 & 24.11 & 0.06 & 53642 & 23.93 & 0.07 & 53647 & 23.78 & 0.07 \\
53670 & 24.35 & 0.06 & 53648 & 23.82 & 0.06 & 53651 & 23.57 & 0.07 \\
53674 & 24.29 & 0.05 & 53651 & 23.76 & 0.05 & 53654 & 23.91 & 0.12 \\
53678 & 24.36 & 0.07 & 53654 & 23.90 & 0.09 & 53666 & 24.28 & 0.20 \\
53700 & 24.27 & 0.01 & 53666 & 23.78 & 0.07 & 53667 & 23.75 & 0.11 \\
53705 & 24.32 & 0.03 & 53670 & 23.81 & 0.04 & 53670 & 23.66 & 0.08 \\
53915 & 25.12 & 0.22 & 53674 & 23.90 & 0.07 & 53674 & 23.68 & 0.06 \\
53916 & 25.81 & 0.27 & 53678 & 24.00 & 0.09 & 53678 & 24.44 & 0.22 \\
53921 & 25.18 & 0.17 & 53681 & 23.76 & 0.08 & 53681 & 23.55 & 0.05 \\
53935 & 25.29 & 0.17 & 53699 & 23.91 & 0.03 & 53699 & 23.64 & 0.04 \\
53939 & 25.32 & 0.16 & 53705 & 23.87 & 0.03 & 53705 & 23.69 & 0.05 \\
53943 & 25.12 & 0.11 & 53709 & 23.86 & 0.09 & 53709 & 23.92 & 0.17 \\
53946 & 25.06 & 0.17 & 53889 & 24.98 & 0.19 & 53909 & 25.06 & 0.29 \\
53963 & 25.10 & 0.17 & 53909 & 25.17 & 0.23 & 53915 & 24.42 & 0.27 \\
53967 & 25.34 & 0.15 & 53915 & 27.40 & 1.07 & 53920 & 25.41 & 0.36 \\
53970 & 25.12 & 0.18 & 53920 & 26.02 & 0.53 & 53935 & 25.78 & 0.49 \\
53975 & 24.99 & 0.10 & 53935 & 25.37 & 0.30 & 53939 & 24.53 & 0.23 \\
53979 & 25.25 & 0.12 & 53939 & 25.44 & 0.24 & 53943 & 25.21 & 0.19 \\
53992 & 25.16 & 0.12 & 53943 & 25.30 & 0.20 & 53963 & 25.56 & 0.35 \\
53996 & 25.49 & 0.15 & 53946 & 25.45 & 0.23 & 53967 & 24.89 & 0.16 \\
54000 & 25.45 & 0.16 & 53963 & 25.30 & 0.24 & 53970 & 25.38 & 0.28 \\
54019 & 26.22 & 0.38 & 53967 & 25.30 & 0.17 & 53975 & 24.92 & 0.20 \\
54023 & 25.22 & 0.12 & 53970 & 25.59 & 0.22 & 53979 & 24.85 & 0.08 \\
54031 & 26.37 & 0.39 & 53975 & 25.49 & 0.18 & 53992 & 25.12 & 0.22 \\
54035 & 25.80 & 0.20 & 53992 & 25.17 & 0.15 & 53996 & 25.43 & 0.32 \\
54053 & 26.23 & 0.38 & 53996 & 25.83 & 0.25 & 54000 & 24.88 & 0.12 \\
54058 & 25.47 & 0.16 & 54000 & 25.95 & 0.31 & 54019 & 26.14 & 0.57 \\
\ldots&\ldots &\ldots& 54019 & 25.05 & 0.16 & 54023 & 24.66 & 0.14 \\
\ldots&\ldots &\ldots& 54023 & 25.56 & 0.24 & 54035 & 25.91 & 0.51 \\
\ldots&\ldots &\ldots& 54031 & 26.12 & 0.54 & 54053 & 24.68 & 0.22 \\
\ldots&\ldots &\ldots& 54035 & 25.67 & 0.26 & 54058 & 25.99 & 0.47 \\
\ldots&\ldots &\ldots& 54053 & 26.48 & 0.68 & \ldots& \ldots&\ldots\\
\ldots&\ldots &\ldots& 54058 & 27.51 & 1.00 & \ldots& \ldots&\ldots\\
\hline
\end{tabular}
\begin{tabular}{l}
\end{tabular}
\caption{SN2213-1745 {\it g'r'i'} magnitudes for 2005 and 2006}
\label{sn224mags}} 
\end{table}

\begin{table}[h!]
\centering 
{\footnotesize
\begin{tabular}{lcccccccc}
\hline 
day (MJD) & {\it g'} & error & day (MJD) & {\it r'} & error & 
day (MJD) & {\it i'} & error\\
\hline
54090 & 26.48 & 0.30 & 54062 & 25.78 & 0.28 & 54061 & 25.71 & 0.30 \\
54096 & 26.10 & 0.24 & 54067 & 26.34 & 0.42 & 54083 & 26.47 & 0.61 \\
54117 & 26.41 & 0.27 & 54083 & 26.36 & 0.56 & 54090 & 25.56 & 0.19 \\
54121 & 26.57 & 0.33 & 54113 & 25.85 & 0.40 & 54095 & 25.85 & 0.33 \\
54124 & 26.46 & 0.33 & 54117 & 25.88 & 0.23 & 54113 & 25.56 & 0.36 \\
54179 & 26.44 & 0.29 & 54122 & 25.98 & 0.26 & 54117 & 25.62 & 0.27 \\
54197 & 26.31 & 0.33 & 54172 & 25.92 & 0.61 & 54120 & 26.05 & 0.31 \\
54202 & 25.61 & 0.18 & 54176 & 26.00 & 0.47 & 54123 & 26.20 & 0.62 \\
54229 & 26.24 & 0.33 & 54183 & 25.85 & 0.31 & 54126 & 26.10 & 0.60 \\
54234 & 25.97 & 0.24 & 54186 & 25.95 & 0.45 & 54140 & 25.99 & 0.42 \\
54237 & 26.05 & 0.23 & 54197 & 26.12 & 0.43 & 54144 & 26.16 & 0.45 \\
54420 & 25.79 & 0.18 & 54201 & 25.45 & 0.23 & 54172 & 25.46 & 0.24 \\
54446 & 25.94 & 0.33 & 54205 & 25.86 & 0.26 & 54176 & 25.98 & 0.59 \\
54452 & 26.18 & 0.28 & 54209 & 25.85 & 0.26 & 54179 & 25.98 & 0.43 \\
54469 & 26.41 & 0.28 & 54213 & 25.96 & 0.37 & 54183 & 25.38 & 0.29 \\
54472 & 25.92 & 0.17 & 54229 & 25.80 & 0.27 & 54186 & 26.02 & 0.59 \\
54476 & 26.25 & 0.22 & 54234 & 25.48 & 0.25 & 54197 & 26.26 & 0.40 \\
54479 & 26.22 & 0.19 & 54237 & 25.30 & 0.17 & 54201 & 25.07 & 0.26 \\
54508 & 26.10 & 0.19 & 54418 & 24.84 & 0.19 & 54205 & 25.49 & 0.21 \\
54510 & 25.69 & 0.23 & 54420 & 25.12 & 0.19 & 54209 & 25.22 & 0.24 \\
54524 & 26.38 & 0.23 & 54445 & 24.78 & 0.17 & 54213 & 25.38 & 0.21 \\
54528 & 26.35 & 0.23 & 54452 & 25.09 & 0.22 & 54229 & 25.15 & 0.18 \\
54531 & 26.31 & 0.22 & 54469 & 25.15 & 0.13 & 54234 & 24.91 & 0.21 \\
54535 & 26.16 & 0.21 & 54472 & 25.16 & 0.13 & 54237 & 25.09 & 0.15 \\
54556 & 26.55 & 0.43 & 54476 & 25.34 & 0.16 & 54418 & 24.66 & 0.16 \\
54558 & 26.11 & 0.19 & 54479 & 25.24 & 0.14 & 54420 & 24.38 & 0.16 \\
54587 & 26.54 & 0.40 & 54481 & 24.81 & 0.17 & 54448 & 24.95 & 0.29 \\
54591 & 26.27 & 0.23 & 54507 & 25.56 & 0.39 & 54452 & 24.87 & 0.20 \\
54615 & 26.01 & 0.21 & 54508 & 25.50 & 0.19 & 54468 & 24.68 & 0.10 \\
\ldots&\ldots &\ldots& 54510 & 25.39 & 0.25 & 54472 & 24.42 & 0.15 \\
\ldots&\ldots &\ldots& 54512 & 25.41 & 0.37 & 54475 & 24.57 & 0.11 \\
\ldots&\ldots &\ldots& 54524 & 25.49 & 0.21 & 54479 & 24.85 & 0.15 \\
\ldots&\ldots &\ldots& 54528 & 25.44 & 0.19 & 54481 & 24.86 & 0.22 \\
\ldots&\ldots &\ldots& 54531 & 25.23 & 0.15 & 54507 & 26.08 & 0.88 \\
\ldots&\ldots &\ldots& 54535 & 25.45 & 0.20 & 54507 & 25.44 & 0.34 \\
\ldots&\ldots &\ldots& 54538 & 25.56 & 0.27 & 54510 & 24.69 & 0.23 \\
\ldots&\ldots &\ldots& 54554 & 25.73 & 0.32 & 54512 & 24.37 & 0.16 \\
\ldots&\ldots &\ldots& 54555 & 25.74 & 0.34 & 54524 & 24.88 & 0.20 \\
\ldots&\ldots &\ldots& 54556 & 25.29 & 0.19 & 54528 & 24.85 & 0.20 \\
\ldots&\ldots &\ldots& 54558 & 25.45 & 0.19 & 54531 & 24.91 & 0.17 \\
\ldots&\ldots &\ldots& 54562 & 25.90 & 0.33 & 54535 & 24.98 & 0.25 \\
\ldots&\ldots &\ldots& 54587 & 25.58 & 0.37 & 54538 & 24.94 & 0.21 \\
\ldots&\ldots &\ldots& 54591 & 25.17 & 0.15 & 54554 & 24.89 & 0.15 \\
\ldots&\ldots &\ldots& 54594 & 25.34 & 0.20 & 54558 & 24.72 & 0.19 \\
\ldots&\ldots &\ldots& 54597 & 25.97 & 0.59 & 54562 & 24.89 & 0.18 \\
\ldots&\ldots &\ldots& 54615 & 25.13 & 0.20 & 54568 & 25.93 & 0.65 \\
\ldots&\ldots &\ldots& 54619 & 25.34 & 0.37 & 54570 & 25.26 & 0.39 \\
\ldots&\ldots &\ldots& \ldots&\ldots &\ldots& 54587 & 24.75 & 0.23 \\
\ldots&\ldots &\ldots& \ldots&\ldots &\ldots& 54591 & 24.65 & 0.19 \\
\ldots&\ldots &\ldots& \ldots&\ldots &\ldots& 54594 & 24.75 & 0.19 \\
\ldots&\ldots &\ldots& \ldots&\ldots &\ldots& 54597 & 24.74 & 0.27 \\
\ldots&\ldots &\ldots& \ldots&\ldots &\ldots& 54614 & 24.56 & 0.13 \\
\ldots&\ldots &\ldots& \ldots&\ldots &\ldots& 54618 & 24.90 & 0.27 \\
\hline
\end{tabular}
\begin{tabular}{l}
\end{tabular}
\caption{SN1000+0216 {\it g'r'i'} magnitudes for 2006 and 2007}
\label{sn279mags}} 
\end{table}

\clearpage

\noindent{\bf REFERENCES} {\it (continued)}